\documentclass[11pt,a4paper]{article}
\pdfoutput=1
\usepackage{jcappub}
\usepackage{bm}
\usepackage{amsmath,amssymb}
\newcommand{\ud}{\mathrm{d}}
\newcommand{\g}{g_{a\gamma}}
\newcommand{\mpl}{m_{\rm pl}}
\newcommand{\xvec}{{\bm{x}}}
\newcommand{\xhat}{\hat{\xvec}}
\newcommand{\yvec}{{\bm{y}}}
\newcommand{\pvec}{{\bm{p}}}
\newcommand{\kvec}{{\bm{k}}}
\newcommand{\Svec}{{\bm{S}}}
\newcommand{\Evec}{{\bm{E}}}
\newcommand{\Bvec}{{\bm{B}}}
\newcommand{\jvec}{{\bm{j}}}
\newcommand{\Jvec}{{\bm{J}}}
\newcommand{\dvec}{{\bm{\nabla}}}
\newcommand{\Cc}{{\mathcal{C}}}

\newcommand{\Beq}{\begin{equation}\begin{aligned}}
\newcommand{\Eeq}{\end{aligned}\end{equation}}

\newcommand{\sech}{\,\textrm{sech}\,}

\title{Dipole Radiation and Beyond from Axion Stars in  Electromagnetic Fields}

\author[a]{Mustafa A. Amin,}
\author[a]{Andrew J. Long,}
\author[a]{Zong-Gang Mou,}
\author[b]{Paul M. Saffin}

\affiliation[a]{Department of Physics and Astronomy, Rice University, Houston, Texas 77005, U.S.A.}
\affiliation[b]{School of Physics and Astronomy, University Park, University of Nottingham,\\ Nottingham NG7 2RD, United Kingdom}

\emailAdd{mustafa.a.amin@rice.edu}
\emailAdd{andrewjlong@rice.edu}
\emailAdd{zm17@rice.edu}
\emailAdd{paul.saffin@nottingham.ac.uk}

\abstract{We investigate the production of photons from coherently oscillating, spatially localized clumps of  axionic fields (oscillons and axion stars) in the presence of external electromagnetic fields. We delineate different qualitative behaviour of the photon luminosity in terms of an effective dimensionless coupling parameter constructed out of the axion-photon coupling, and field amplitude, oscillation frequency and radius of the axion star. For small values of this dimensionless coupling, we provide a general analytic formula for the dipole radiation field and the photon luminosity per solid angle, including a strong dependence on the radius of the configuration. For moderate to large coupling, we report on a non-monotonic behavior of the luminosity with the coupling strength in the presence of external magnetic fields. After an initial rise in luminosity with the coupling strength, we see a suppression (by an order of magnitude or more compared to the dipole radiation approximation) at moderately large coupling. At sufficiently large coupling, we find a transition to a regime of exponential growth of the luminosity due to parametric resonance. We carry out 3+1 dimensional lattice simulations of axion electrodynamics, at small and large coupling, including non-perturbative effects of parametric resonance as well as backreaction effects when necessary. We also discuss medium (plasma) effects that lead to resonant axion to photon conversion, relevance of the coherence of the soliton, and implications of our results in astrophysical and cosmological settings.
}

\begin{document}

\maketitle

\section{Introduction}\label{sec:intro}

Axions and axion-like particles provide excellent dark matter candidates \cite{Preskill:1982cy,Abbott:1982af,Dine:1982ah,Kim:1986ax,Cheng:1987gp,Turner:1989vc,Raffelt:1990yz,Duffy:2009ig,Ringwald:2014vqa}, as well as candidates for driving inflation \cite{Freese:1990rb,Choi:1999xn,Silverstein:2008sg,Pajer:2013fsa} and, perhaps, even present day acceleration \cite{Kim:2002tq,Chacko:2004ky,Hall:2005xb,Barbieri:2005gj}. While originally motivated as a solution to the strong CP problem \cite{Peccei:1977hh,Weinberg:1977ma,Wilczek:1977pj}, they are ubiquitous in many high energy physics theories \cite{Banks:2003sx,Denef:2004dm,Svrcek:2006yi,Arvanitaki:2009fg}. A variety of experimental efforts are underway to detect axions and axion-like particles (ALPs) in the laboratory \cite{Aprile:2020tmw,Salemi:2021gck,Jiang:2021dby,Devlin:2021fpq} and through their unique astrophysical and cosmological signatures \cite{Ayala:2014pea,Marsh:2015xka,Poulin:2018dzj,Reynolds:2019uqt, Arvanitaki:2019rax,Sigl:2019pmj,Ferreira:2020fam,Depta:2020wmr,Athron:2020maw,Hui:2021tkt}. Many of these searches rely upon a coupling of the axion field $\phi(\xvec,t)$ to electromagnetism via the interaction $\g \phi \bm{E} \cdot \bm{B}$. Constraints on this coupling depend on the axion's mass, and they are typically at the level of $\g \lesssim \mathrm{few} \times 10^{-11} \ \mathrm{GeV}^{-1}$ for a light dark matter axion.  

Since the $\g\phi\bm{E}\cdot\bm{B}$ interaction between axions and electromagnetism is expected to be very weak, one might seek to compensate the tiny coupling $\g$ by searching for signatures in systems with a strong electromagnetic field and/or a large axion field amplitude.
Moreover, if $\phi$ is oscillating it can also have an enhanced effect through resonances \cite{Arza:2020eik,Hertzberg:2020dbk,Amin:2020vja}. 
With these considerations, it is natural to explore the impact of axion stars in strong electromagnetic fields.  
Axion stars are large amplitude, spatially localized and oscillating $\phi$ field configurations (also known as scalar solitons, oscillons, axitons etc. \cite{Kaup:1968,Bogolyubsky:1976pw,Bogolyubsky:1976yu,Seidel:1991zh,Kolb:1993zz,Gleiser:1993pt,Copeland:1995fq,Kasuya:2002zs,Amin:2013ika,Amin:2010jq,Visinelli:2017ooc,Hertzberg:2018lmt,Guerra:2019srj,Eby:2019ntd}).
Such exploration is the main purpose of this paper. 
We aim to understand how the interplay between the axion-photon coupling strength, parameters defining solitons, and electromagnetic fields influences the production of electromagnetic radiation from such solitons.

The study of axions in astrophysical and cosmological electromagnetic fields has a long history~\cite{Raffelt:2006cw,Fortin:2021sst}.  
Some of the strongest constraints on the coupling of axions to matter come from considering the production of axions in the hot and dense stellar interiors.  
Alternatively, if axions were produced in the early universe and survive today as dark matter, then the flux of these cold axions onto magnetized compact stars could result in a distinctive radio emission \cite{Pshirkov:2007st,Sigl:2017sew,Huang:2018lxq,Hook:2018iia,Caputo:2018ljp,Battye:2019aco,Leroy:2019ghm,Foster:2020pgt}.  
As much as an $\mathcal{O}(1)$ fraction of the axion dark matter could be in the form of axion stars, and therefore it is also important to develop strategies for detecting the encounter of axion stars with magnetized compact stars \cite{Tkachev:2014dpa,Iwazaki:2014wka,Raby:2016deh,Pshirkov:2016bjr,Iwazaki:2017rtb,Eby:2017xaw,Bai:2017feq,Iwazaki:2018squ,Buckley:2020fmh,Prabhu:2020yif,Edwards:2020afl,Wang:2021wae,Nurmi:2021xds}. Furthermore, the collision of and collapse of axion stars can amplify even small fluctuations in the electromagnetic fields \cite{Tkachev:2014dpa, Hertzberg:2018lmt,Levkov:2018kau,Amin:2020vja}.

In the work being presented here, we consider the coherent emission of electromagnetic radiation from an axion star in an electromagnetic field. Because of the high occupation number of the axions in the solitons, it is natural to treat the axion field classically.  
We calculate the spectrum and luminosity of the resultant electromagnetic radiation using both analytical techniques and 3+1 dimensional lattice simulations. Based on Floquet analysis, we argue that different qualitative behaviour of the electromagnetic radiation is determined by a dimensionless effective coupling parameter $\mathcal{C}\sim (\g \varphi_0)(\omega R)$, where $\varphi_0$ is the field amplitude, $\omega$ and $R$ are the oscillation frequency and radius of the axion star respectively.

In the small effective coupling $\mathcal{C}\ll 1$ regime, the analytical analysis is based on the observation that an axion star in an external electromagnetic field develops a charge and current dipole, which oscillates in time and produces dipole radiation. In the absence of resonance effects, we find that the signal has a strong dependence on the axion star's radius and oscillation frequency, which leads to a suppression (that goes like the Fourier transform of the spatial profile) at large $\omega R > \mathcal{O}(1)$. This understanding leads us to focus our attention on compact axion-star configurations and oscillons. Our dipole approximation is validated  with numerical simulations of the axion electrodynamics on a $3+1$ dimensional lattice. 

Floquet analysis and lattice simulations also allow us to study the regime of moderate to large $\mathcal{C}\sim 1$,  where perturbative analytical results are difficult to obtain. We are able to capture a non-trivial transition from a steady photon production rate to an explosive (exponential) one as we vary the coupling strength and axion field configuration. We find qualitative and quantitative differences between the photon production rate in the presence of external electric and magnetic fields (including significant suppression of the radiated power at moderate coupling). We also analyse the backreaction of photons on the axion configuration when necessary.

Most earlier work on axion stars in astrophysical magnetic fields relies on a `resonant' axion-to-photon conversion, when the plasma frequency approximately matches the energy of the axion particles (see, for example, \cite{Raffelt:1990yz}). While our simulations do not include the effects of a plasma,  we are able to incorporate this resonant conversion in our calculation analytically in the small coupling regime. We also comment on the relevance of a coherent solitonic configuration compared to an incoherent collection of dipoles, as well as the connection of our results to the well-known quantum mechanical calculation related to the axion-photon conversion probability (see, for example, \cite{Raffelt:2006cw}).

The remainder of this article is organised as follows: 
In Sec.~\ref{sec:Axion_electrodynamics} we briefly introduce the model of interest, namely axion electrodynamics, and in Sec.~\ref{sec:axion_stars} we introduce axion stars.
In Secs.~\ref{sec:Analytic} and \ref{sec:lattice} we employ analytical and numerical techniques to calculate the spectrum of electromagnetic radiation that arises from an axion star in an external electromagnetic field. 
In Sec. \ref{sec:medium_effects} we comment on a few supplemental topics such as finite density and coherence effects, and in Sec.~\ref{sec:observability} we discuss several possible observational signatures.
Finally, we summarize and conclude in  Sec.~\ref{sec:conclusion}. We include an Appendix \ref{sec:AppendixA} with details of the dipole radiation calculation.

\section{Axion electrodynamics}
\label{sec:Axion_electrodynamics}

Our system consists of a real valued, pseudo-scalar field $\phi$ coupled to the electromagnetic field. The action for our system is given by
\Beq
\label{eq:action_c}
S=\int \! \mathrm{d}^4 x\left[-\frac{1}{2}\partial_\mu\phi\partial^\mu\phi-V(\phi)-\frac{1}{4}F_{\mu\nu}F^{\mu\nu}-\frac{\g}{4}\phi F_{\mu\nu}\tilde{F}^{\mu\nu}\right]\,,
\Eeq
where we adopt the $-+++$ signature of the metric. The electromagnetic field-strength tensor, and its dual are:
\Beq
F_{\mu\nu}=\partial_\mu A_\nu-\partial_\nu A_\mu\,,\qquad\tilde{F}^{\mu\nu}=\frac{1}{2}\epsilon^{\mu\nu\rho\sigma}F_{\rho\sigma},
\Eeq
where $\epsilon^{0123}=1$. The equations of motion for the axion and the gauge fields are given by 
\Beq
&\partial_\mu\partial^\mu\phi-\partial_\phi V=\frac{\g}{4} F_{\mu\nu}\tilde{F}^{\mu\nu}\,,\\
&\partial_\mu F^{\mu\nu}=-j^\nu\,,\qquad \partial_\mu \tilde{F}^{\mu\nu}=0\,,\\
\Eeq
where
\Beq
j^\nu\equiv \g \partial_\mu\phi \tilde{F}^{\mu\nu}\,.
\Eeq
Note that $\partial_\mu j^{\mu}=0$, and that we have assumed that there are no free currents or charges in our system. The above four-current arises from axion-electromagnetic interactions.  

We define electric and magnetic fields in the usual way
\Beq
E_i=F_{i0}\qquad \textrm{and}\qquad B_i=(1/2)\epsilon_{ijk}F^{jk}\,,
\Eeq
with $\epsilon_{ijk}=\epsilon^{ijk}$. The coupled Klein-Gordon and Maxwell equations are then given by~\cite{Sikivie:1983ip}
\Beq
\label{eq:KGMax}
&\ddot{\phi}-\nabla^2\phi+\partial_\phi V=\g \bm{E}\cdot\bm{B}\,,\\
&\dot{\bm{E}}=\nabla \times \bm{B}-\g\left(\dot{\phi}\bm{B}+\nabla\phi\times \bm{E}\right)\,,\\
&\dot{\bm{B}}=-\nabla\times \bm{E}\,,\\
&\nabla \cdot \bm{E}=-\g\nabla\phi \cdot{\bm{B}}\,,\\
&\nabla \cdot \bm{B}=0\,.
\Eeq
Note that the effective charge and current densities are
\Beq
\label{eq:DensityCurrent}
\rho= -\g\nabla\phi \cdot\bm{B}\,\qquad\textrm{and}\qquad {\bm{ J}}= \g\left(\dot{\phi}\bm{B}+\nabla\phi\times \bm{E}\right)\,.
\Eeq
In the above equations, we have ignored gravitational interactions. If one wishes to include weak-field gravity (gravitational potential $|\Psi|\ll 1$), the substitution $\partial_\phi V\rightarrow (1+2\Psi)\partial_\phi V$ in the equation of motion for the scalar field captures the most relevant gravitational contributions. Moreover, we would need to include a Poisson equation $\nabla^2\Psi=(1/2\mpl^2)\rho_\phi$ where $\rho_\phi$ is the density of the axion field to close the system. This prescription allows certain gravity-supported scalar field configurations to exist, but ignores gravitational effects (such as redshifts) in the dynamics of electromagnetic fields and also ignores the contribution of electromagnetic fields in determining the gravitational potential.\footnote{We are also assuming $\g$ is sufficiently small here, and the electromagnetic fields are the subdominant contribution to the total energy density of the system.}

\section{Compact axion stars in constant electromagnetic fields}
\label{sec:axion_stars}

We are interested in electromagnetic radiation generated by a spatially localized, spherically symmetric, coherently oscillating axion field configuration of the approximate form
\Beq
\label{eq:oscillon}
\phi(t,r)\approx \varphi(r)\cos(\omega t)\,.
\Eeq
Such solutions of the nonlinear Klein-Gordon equation (with and without gravity), which we generically refer to as solitons, are a result of a balance between the tendency of the field configurations to disperse and (i) attractive self-interactions in the potential $V(\phi)$  and/or (ii) gravitational interactions. 

The detailed form of $\varphi(r)$ depends on the potential $V(\phi)$ as well as $\omega$. For most of our purposes, we use an ansatz of the form $\varphi(r)=\varphi_0 \sech(r/R)$ so that
\Beq
\label{eq:sech}
\phi(t,r)=\varphi_0 \sech(r/R) \cos\omega t\,.
\Eeq
The above form is motivated by the fact that it has the correct large distance  behavior: $\sim e^{-r/R}$, with $R\sim 1/\sqrt{m^2-\omega^2}$ where $m>\omega$. Note that there is also polynomial dependence of the profile on $r$ at large radii multiplying the exponential, which are ignoring here \cite{Amin:2010dc,Schiappacasse:2017ham}.  Typically, $\omega$ is not too different from $m$, however, $\varphi_0$ and $R$ can vary significantly for small changes in $\omega$ close to $m$. In a typical scenario, $\varphi_0$, $R$ and $\omega$ are not independent. Usually we are free to chose only one, and even that has constraints from stability analyses. 

To understand what to expect for $\varphi_0$ and $R$, we consider two relevant cases below.

\subsection{Self-interaction supported solitons}\label{sec:self_int}
\begin{figure}[t]
\begin{center}
\includegraphics[width=1\textwidth]{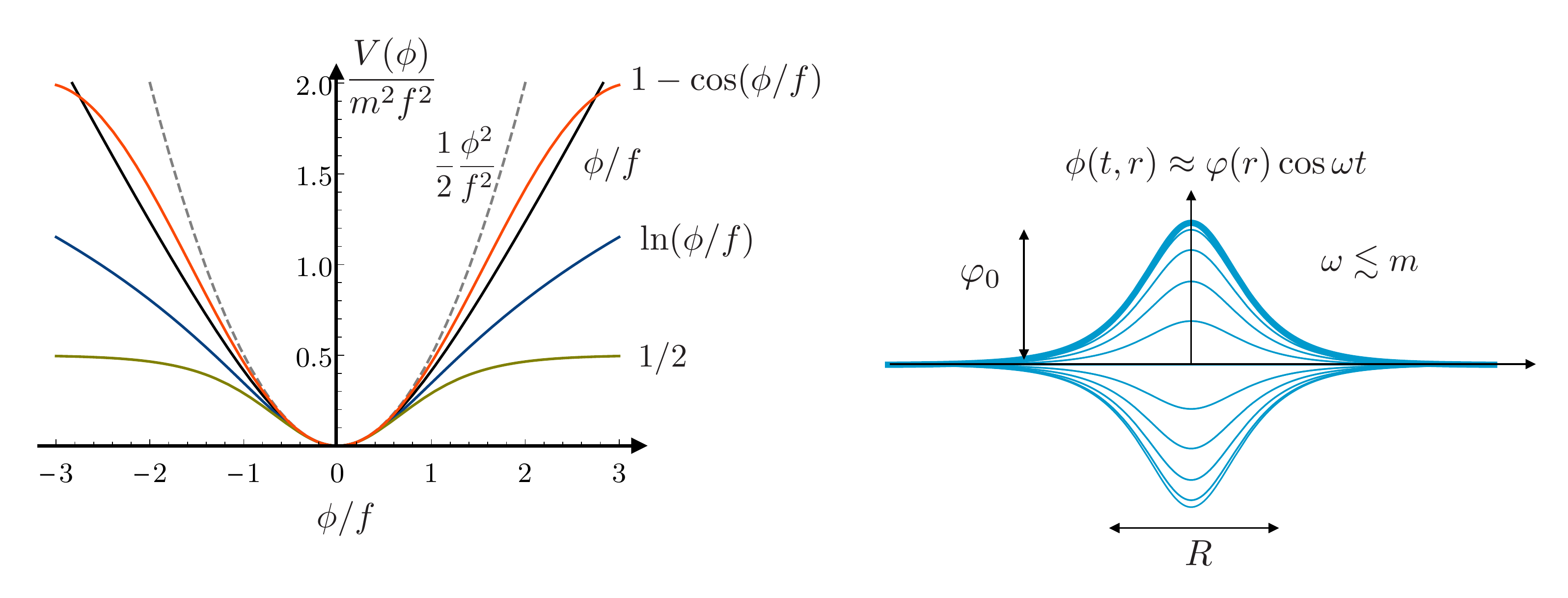}
\end{center}
\caption{
Left: The scalar field potentials that support solitons. For the quadratic potential and cosine potential, gravity is essential for supporting long-lived solitons, whereas the ``flattened" potentials can support solitons without gravity, but typically require amplitudes $\sim f$. For any potential where solitons have a small amplitude compared to $f$, gravity is essential for long-lived stable solitons. Right: A schematic representation of a solitons. Dilute solitons have $\varphi_0\ll f$ and $R\gg m^{-1}$. Dense solitons have $\varphi_0\sim f$ and $R\sim {\rm few}\times m^{-1}$. The frequency is always $\omega\lesssim m$.} 
\label{fig:potential}
\end{figure}

For potentials that have $V(\phi)\approx (1/2)m^2\phi^2$ where $\phi\ll f$ and $V(\phi)\propto \phi^{\alpha<2}$ where $\phi\gg f$ (see Fig.~\ref{fig:potential}), exceptionally long lived spatially localized configurations of the above form exist, and are called oscillons (setting $\g\rightarrow 0$ for the moment). Typically, for very long-lived oscillons, we have a field amplitude $\varphi_0\sim f$, a spatial width $R\sim {\rm few}\times m^{-1}$ and an oscillation frequency $\omega\lesssim m$ \cite{Zhang:2020bec}. In detail, there is a one parameter family of long lived configurations for a given potential $V(\phi)$. Moreover, typically the solution also includes higher frequencies and a very small radiating tail (scalar radiation \cite{Fodor:2019ftc}).

Because of the scalar radiation, oscillons are not perfectly stable, and they exhaust their energy on a time scale $\tau$.  This lifetime depends sensitively on the scalar potential $V(\phi)$.  For example, a cosine potential leads to a lifetime $\tau \sim 10^3 m^{-1}$, whereas some other potentials shown in Fig.~\ref{fig:potential} give much longer lifetimes $\tau \gtrsim 10^{12} m^{-1}$ \cite{Olle:2019kbo,Zhang:2020bec,Zhang:2020ntm}. The formation of such oscillons from cosmological initial conditions (especially in the early universe) has been explored in detail before \cite{Kolb:1993hw,Amin:2011hj,Gleiser:2011xj,Lozanov:2017hjm,Hong:2017ooe}, and typically happens when $H\sim m$. An almost homogeneous, oscillating condensate naturally fragments into oscillons. As a result, compared to the $H^{-1}$ at the time of formation, oscillons can be exceptionally long lived, and have important cosmological implications \cite{Lozanov:2019ylm,Olle:2019kbo,Amin:2019ums,Zhou:2013tsa,Liu:2017hua}. Using similar arguments, oscillons in ultra-light axions might potentially survive until today \cite{Kawasaki:2019czd,Olle:2019kbo,Ibe:2019vyo}. Furthermore, oscillons appear to be attractors in the space of solutions \cite{Gleiser:2009ys}, and might also nucleate inside dark matter halos \cite{Levkov:2018kau}, or even near black-holes \cite{Hertzberg:2020hsz}, although most of these analyses are done in the context of gravitationally supported solitons so far. 

 If we are interested in a population of oscillons in the contemporary universe which have a primordial origin, their lifetimes will likely be short compared to $H_0^{-1}\sim 10^{33} \ \mathrm{eV}^{-1}$ (unless $m \lesssim 10^{-21} \ \mathrm{eV}$). While challenging, claims exist in the literature for oscillons that have lifetimes comparable to the present age of the universe \cite{Gleiser:2019rvw,Olle:2020qqy}. As discussed earlier, late universe formation mechanisms can also ameliorate this problem (also see \cite{Arvanitaki:2019rax}). The detailed investigation of oscillon production and population is not considered in this paper; we simply take these objects to exist and examine their consequences due to encounters with external electromagnetic fields.

\subsection{Gravitationally supported non-relativistic solitons}
It is also possible to obtain solutions of the form in eq.~\eqref{eq:oscillon} for $V(\phi)= (1/2)m^2\phi^2$ (ie. without nonlinearities in the potential) as long as we now allow for gravitational interactions \cite{Kaup:1968,Seidel:1991zh}. Such configurations are sometimes referred to as oscillatons \cite{UrenaLopez:2001tw}. Such oscillatons can be compact, with $R\sim 10 m^{-1}$, with an amplitude $\varphi_0\sim 0.1 
\mpl$ \cite{Kaup:1968,Alcubierre:2003sx}. For some formation mechanisms, see \cite{Eggemeier:2019jsu,Widdicombe:2018oeo,Levkov:2018kau,Schwabe:2016rze,Chen:2020cef}.

Less compact configurations can also exist if the amplitude of the field is not so large, and $R\gg m^{-1}$ -- they are referred to as dilute axion stars \cite{Visinelli:2017ooc,Eby:2019ntd}. In this regime the central field amplitude $\varphi_0\propto 1/R^2$, the frequency $\omega \approx m$, and the radius $R\sim 1/\sqrt{m^2-\omega^2}$. Such dilute configurations are well described by a Schr\"{o}dinger-Poisson system (with a conserved particle number), and are often the focus in fuzzy dark matter studies \cite{Hu:2000ke,Hui:2016ltb,Olle:2019kbo}. Dilute axion stars have the benefit of being cosmologically long-lived. However, as we will see, electromagnetic radiation from dilute axion stars in external electric and magnetic fields is heavily suppressed. As a result, most of our focus will be on the dense, smaller radius solitons.\footnote{Solitons can also form with gravitational and repulsive self-interactions. See, for example, \cite{Croon:2018ybs}.}

\section{Analytic calculation of electromagnetic radiation}
\label{sec:Analytic}

In this section, we calculate the electromagnetic radiation generated by spatially localized, coherently oscillating axion configurations (solitons) discussed in the previous section. In the presence of external electromagnetic fields, such configurations can be effectively thought of as time-dependent charge densities and currents that produce electromagnetic radiation. We provide analytic results for the produced radiation at leading order in the coupling $\g$, and discuss difficulties with going beyond the leading order analytically. We also discuss the expected non-perturbative (in the coupling) results in general terms. 

The first-order Maxwell equations \eqref{eq:KGMax} can be rearranged into the following differential equations:
\begin{align}
\label{eq:EBwave}
\ddot {\bm{E}} - {\bm{\nabla}}^2 {\bm{E}} = - {\bm{\nabla}}\rho -\dot {\bm{J}}\,,\qquad
\ddot {\bm{B}} - {\bm{\nabla}}^2 {\bm{B}} =  {\bm{\nabla}}\times {\bm{J}}\,.
\end{align}
The 4-current $(\rho,{\bm{J}})$ defined in \eqref{eq:DensityCurrent} is spatially localized because the axion field configuration $\phi$ given by eq.~\eqref{eq:oscillon} is spatially localized. Note that $(\rho,{\bm{J}})$ depend on $\phi$ as well as the $\bm{E}$ and $\bm{B}$ via eq.~\eqref{eq:DensityCurrent}. Beyond the spatial extent of the axion stars, both ${\bm{E}}$ and ${\bm{B}}$ propagate like free waves. 

\subsection{Floquet analysis}
Because the system is linear in $\bm{E}$ and $\bm{B}$ fields, and we assume $\phi$ to be periodic in time, we expect the solutions to obey Floquet's Theorem \cite{HillBook,dif}. That is, the solutions are either bounded and periodic, or have exponential growth in time. However calculating Floquet exponents ($\mu$), or explicit solutions is a tall order because of the large number of coupled degrees of freedom associated with each spatial point (formally infinite, and usually a rather large number in discretized three dimensions). Equivalently, the modes in Fourier space are coupled because of the spatial variation in $\phi$.\footnote{The number of Floquet exponents is equal to the dimensionality of phase space for the system. For a system with $N^3$ Fourier modes, there would be $2N^3$ Floquet exponents. For a coupled system of Fourier modes (ie. inhomogeneous background), each Floquet exponent does not correspond to a single Fourier mode, but a linear combination of modes. Note that Floquet exponents are complex in general. When they have a non-zero real part, we can get exponential solutions in time. When we refer to Floquet exponents from here onwards, we are referring to the real part.}

While the explicit calculation of the Floquet exponents is non-trivial, we can get a physical understanding of their scaling with parameters and the parametric boundary between bounded and unbounded solutions as follows. For the homogeneous axion field with amplitude $\varphi_0$ and oscillating harmonically with a frequency $\omega$, the electromagnetic fields are always unstable, with the $k\approx \omega/2$ electromagnetic field modes growing as $e^{\mu_{\rm hom}t}$ where $\mu_{\rm hom}\approx \g\varphi_0 \omega/4$ at least when $\g\varphi_0$ is not too large \cite{Hertzberg:2018zte} (for larger amplitudes, it is model dependent \cite{Amin:2020vja}). In contrast, for the localized soliton configuration, we expect a threshold value of the coupling $\g \varphi_0$ for which we get exponentially growing solutions. The parameter $\varphi_0$ should now also be thought of as the central amplitude of the soliton. The threshold can be determined by comparing $\mu_{\rm hom}^{-1}$ to the width of the soliton $R$ \cite{Hertzberg:2010yz,Hertzberg:2018zte,Amin:2020vja}. Essentially, if the produced photons can escape the system quickly enough (ie. $R$ is small enough), they do not lead to exponential growth due to parametric resonance (equivalently, Bose-enhancement). This motivates the definition of a dimensionless effective coupling
\Beq
\label{eq:EffCoup_def}
\mathcal{C}\equiv \frac{R}{\mu_{\rm hom}^{-1}}\approx \frac{1}{4}\g \varphi_0 \omega R.
\Eeq
In terms of this effective coupling:
\Beq
\label{eq:EffCoup_crit}
&\mathcal{C}\ll 1\,\longrightarrow\, \textrm{bounded periodic solutions, steady radiated power\,,}\\
&\mathcal{C}\gtrsim 1\,\longrightarrow
\,\textrm{unbounded exponential solutions and radiated power}\,.\\
\Eeq
We remind the reader that $\mathcal{C}$ is independent of background electromagnetic fields. Note that for $\mathcal{C}> \mathcal{C}_{\rm crit}\sim 1$, the power in radiated electromagnetic fields 
\Beq
P_\gamma\propto e^{2\mu_{\rm eff}t}\,\qquad\textrm{where} \qquad\mu_{\rm eff}\propto\g \varphi_0\omega\,,
\Eeq
In Sec.~\ref{sec:lattice}, we will confirm this behaviour, and provide the numerical coefficient in front of this expression for $\mu_{\rm eff}$ based on a specific soliton profile.

We remind the reader that soliton configurations do not allow us to specify $\varphi_0$, $\omega$ and $R$ independently. For example, dilute and gravitationally supported solitons have $\varphi_0\propto R^{-2}$.  For dense, self-interaction supported axion stars/oscillons, $\varphi_0\sim f$. For the dilute case, we have $\omega R\gg 1$, so we can get $\mathcal{C}\sim 1$ for $\g\varphi_0\ll 1$. For the dense case, we typically have $R\sim {\rm few}\times m^{-1}$, so we can get $\mathcal{C}\sim 1$ with $\g \varphi_0\sim 1$. The $\mathcal{C}\ll 1$ can be achieved, for example, by simply making $\g$ smaller in each case. 

Before moving on to a quantitative analytical analysis, we briefly discuss the connection of $\Cc\ll1$ and $\Cc\gtrsim 1$ regimes with effective field theory (EFT) considerations. The action in Eq.~\eqref{eq:action_c} represents the leading operators in an EFT with cutoff $\Lambda \sim \g^{-1}$ describing axion-photon interactions.\footnote{If the axion-photon interaction is loop-induced, such as for models of the QCD axion, then one expects $\g \approx \alpha / 2 \pi f \sim 10^{-3} / f$.  However in this work we take a more general approach by treating $\g$ and $f$ as independent parameters where $f$ enters as a scale in the axion potential.}  
The EFT also contains sub-leading operators that are suppressed by additional powers of the cutoff, e.g. $\mathcal{L}_\mathrm{sub} \supset c_\mathrm{sub} \, \g^2 \phi^2 F^2$ or $c_\mathrm{sub} \, \g^3 \Box \phi F\tilde{F}$. Validity of the EFT requires the sub-leading operators to be negligible. As discussed above, it is possible to have $\g \varphi_0\ll 1$ to get $\Cc\ll 1$. For dilute axion stars, $\Cc\sim 1$ can be obtained for $\g \varphi_0\ll 1$ also. However, for $\Cc\sim 1$ in the dense case, we need $\g \varphi_0\sim 1$, which threatens to break the EFT if higher-order operators are only suppressed by additional powers of $\g \varphi_0$.  
Even in this case, the EFT can remain reliable even for $\g \varphi_0 \sim 1$ if the numerical coefficient of the higher-order operators is small, e.g. $c_\mathrm{sub} \ll 1$. 
For some theoretical work on models with a large axion-photon coupling, see \cite{Choi:2014rja,Choi:2015fiu,Kaplan:2015fuy,Farina:2016tgd,Agrawal:2018mkd,Daido:2018dmu,Dror:2020zru}. 

\subsection{Perturbative analysis}
With the expectation of bounded solutions for $\mathcal{C}\ll 1$, we pursue an analytic treatment in the limit of small $\g\varphi_0$. With this small parameter in mind, we expand the fields, densities and currents as follows:
\begin{align}
\label{eq:pertur}
&{\bm{E}} ={\bm{E}}_{(0)} +{\bm{E}}_{(1)} +{\bm{E}}_{(2)} +\cdots
,
\quad
&&{\bm{B}} ={\bm{B}}_{(0)} +{\bm{B}}_{(1)} +{\bm{B}}_{(2)} +\cdots
,
\\
&{\rho} ={\rho}_{(0)} +{\rho}_{(1)} +{\rho}_{(2)} +\cdots
,
\quad
&&{\bm{J}} ={\bm{J}}_{(0)} +{\bm{J}}_{(1)} +{\bm{J}}_{(2)} +\cdots
.
\end{align}
Here we use the subscript ${}_{(n)}$ to denote the terms containing $n$-th power of $\g \varphi_0$.

At the lowest order, the ${\bm{E}}_{(0)}$ and ${\bm{B}}_{(0)}$ stand for the electric and magnetic backgrounds and are sourced by $(\rho^{(0)},{\bm{J}}^{(0)})$ which are independent of the axion field configuration. For example such background fields could be the fields in the magnetosphere of a neutron star or in the intergalactic medium. To make the physics more transparent, we will consider spatio-temporally constant background electromagnetic  fields which we denote by  
\Beq
{\bm{E}}_{(0)}=\bar{\bm{E}}\,,\qquad \textrm{and}\qquad {\bm{B}}_{(0)}=\bar{\bm{B}}\,.
\Eeq We are essentially assuming that the spatial extent of the axion star is much smaller than the coherence length of the background fields, and that the time variation of the background fields is slow compared to the time that configuration spends in the given volume of the fields.

\begin{figure}[t]
\begin{center}
\includegraphics[width=0.7\textwidth]{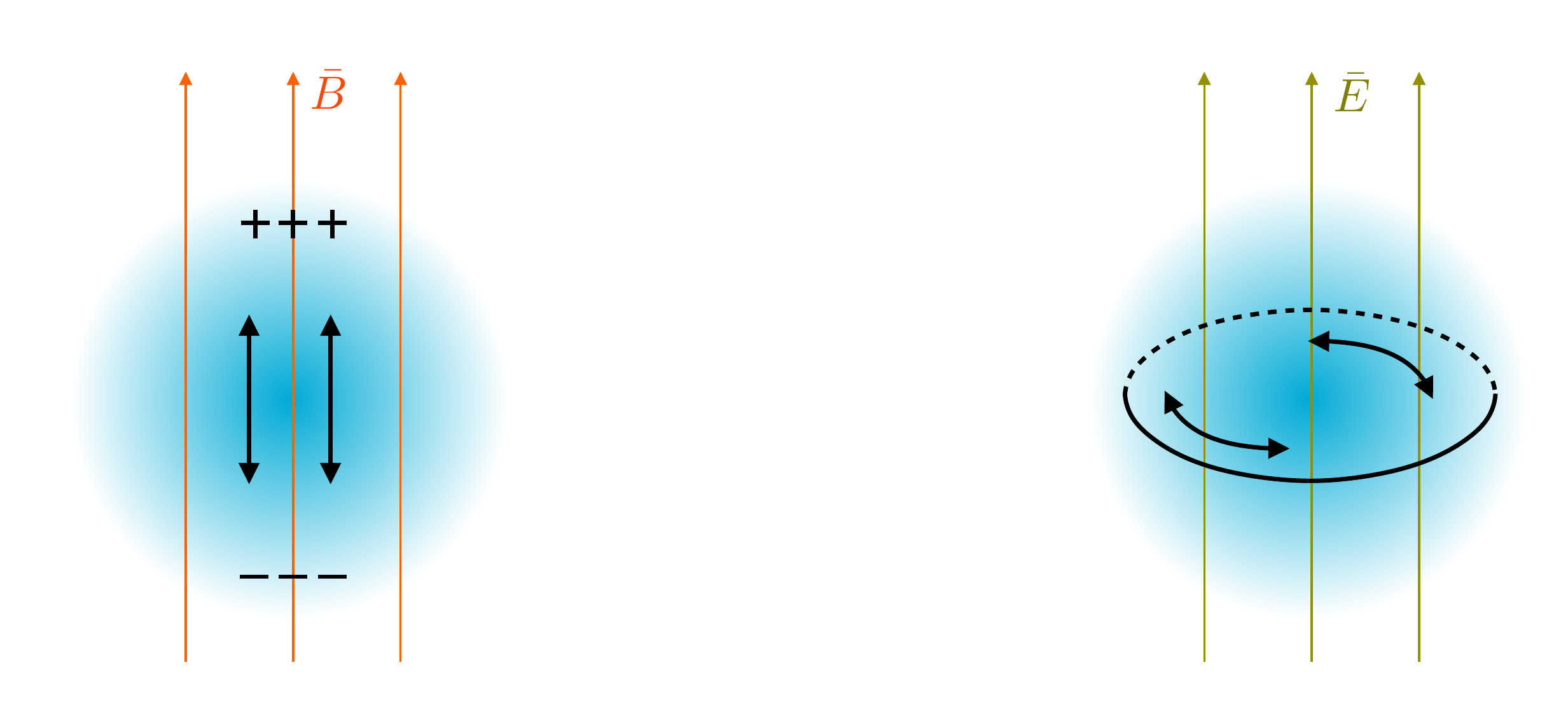}
\end{center}
\caption{
The effective charge and current density (dipoles) induced by the presence of a soliton in an external electromagnetic field background.  The left image shows a charge dipole aligned with the external magnetic field, and the right image shows a current dipole in a plane normal to the external electric field.  The charge density and current density oscillate in time, generating dipole radiation.} 
\label{fig:charge}
\end{figure}
\subsubsection{Leading order in $\g\varphi_0$: dipole radiation}
\label{sec:dipole}
At leading order in the coupling $\g\varphi_0$, we have
\begin{align}
\label{eq:Ewave1}
\ddot {\bm{E}}_{(1)} - {\bm{\nabla}}^2 {\bm{E}}_{(1)} &= - {\bm{\nabla}}\rho_{(1)} -\dot {\bm{J}}_{(1)}
,
\\
\label{eq:Bwave1}
\ddot {\bm{B}}_{(1)} - {\bm{\nabla}}^2 {\bm{B}}_{(1)} &=  {\bm{\nabla}}\times {\bm{J}}_{(1)}.
\end{align}
At this order in $\g\varphi_0$, the background electromagnetic fields along with the axion configuration $\phi(t,\xvec)=\varphi(r)\cos\omega t$ induce an effective charge and current density:
\begin{align}
\label{eq:harmonicsource}
 &\rho_{(1)}(t,\xvec) 
= {\rm Re}\left[ \varrho_{(1)}(\xvec)  e^{-i\omega t}\right]
,
\quad
&&\bm{J}_{(1)}(t,\xvec)   = {\rm Re}\left[ \bm{j}_{(1)}(\xvec)  e^{-i\omega t}\right]
,\\
\hspace{-0.4cm}{\rm with}\quad
 &\varrho_{(1)}(\xvec) 
=  -g_{a\gamma}  {\bm{\nabla}}\varphi(r) \cdot \bar{\bm{B}}
,
\quad
&&\bm{j}_{(1)}(\xvec)   =  -i\omega g_{a\gamma}  \varphi(r) \bar{\bm{B}} +g_{a\gamma}  {\bm{\nabla}}\varphi(r) \times \bar{\bm{E}}.
\label{eq:current}
\end{align}
Due to the spatial derivative acting on $\varphi$ along the direction of $\bar{\bm{B}}$ field, the positive and the negative charges are distributed separately along the $\bar{\bm{B}}$ field axis like a dipole (see left panel in Fig.~\ref{fig:charge}).
And with its oscillating nature of the axion configuration, such an oscillating dipole will lead to dipolar electromagnetic radiation. A constant $\bar{\bm{E}}$ field results in an oscillating azimuthal current, which also results in dipolar radiation (see right panel in Fig.~\ref{fig:charge}).

\begin{figure}[t]
\begin{center}
\includegraphics[width=1.0\textwidth,trim=0cm 0cm 0cm 0cm, clip]{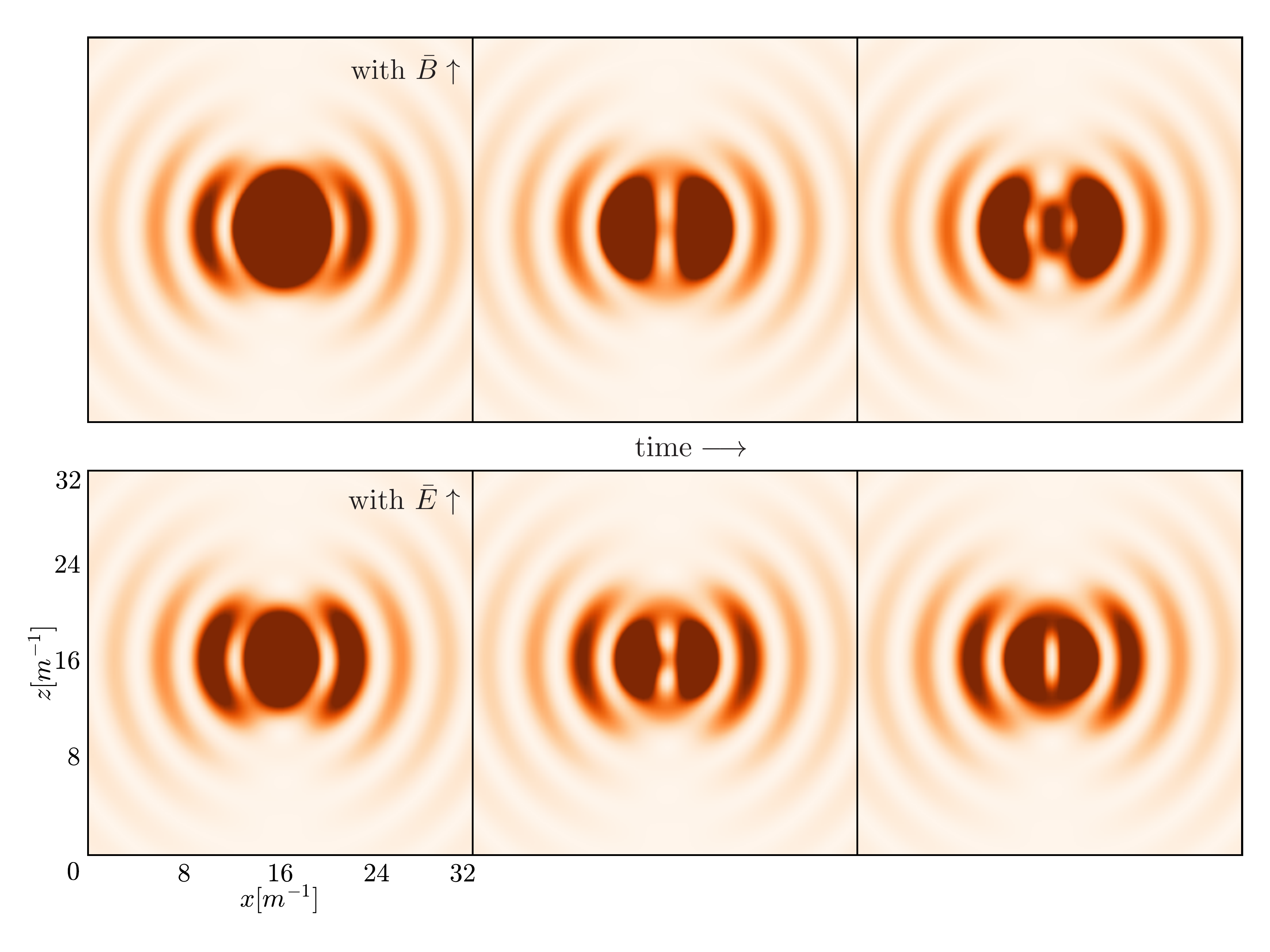}
\end{center}
\caption{(Top panels) Energy density of the emitted electromagnetic radiation from an oscillating electric dipole created by an axion star at rest in a background magnetic field. (Bottom panels) Dipole radiation from an  oscillating magnetic dipole created by an axion star at rest in a background electric field. The magnetic and electric fields point in the $z$ direction. The colors represent electromagnetic energy density $\epsilon_\gamma=(1/2)(\bm{E}^2+\bm{B}^2)$ after subtracting the background fields. Darker colors represent higher energy densities. For visual clarity, we have allowed for colors to be saturated in the densest regions.}
\label{fig:dipoleRad}
\end{figure}

It is a standard textbook problem to compute the excited electric and magnetic fields caused by the harmonic, spatially localized sources of the form \eqref{eq:harmonicsource}, as well as the associated Poynting flux $\Svec_{(2)}\equiv \bm{E}_{(1)}\times \bm{B}_{(1)}$ and power emitted per unit solid angle. See for example \cite{Jackson:100964,Vilenkin:2000jqa}. We review some of the relevant details of the derivation in Appendix \ref{sec:AppendixA}. Here, we directly write down the solution for the flux below. 
At a position $\xvec$ far from the source, and at sufficiently late times, the power per unit solid angle $dP^\gamma_{(2)}/d\Omega=|\xvec|^2\xhat\cdot\Svec_{(2)}$, is given by
\begin{align}
    \label{eq:dipoleGen}
    \frac{dP^\gamma_{(2)}}{d\Omega} & =
    \frac{\omega^2}{32\pi^2} 
    \Bigg( \!\!
    -|\tilde{\varrho}_{(1)} (\kvec)|^2
    +|\tilde{\jvec}_{(1)}(\kvec)|^2
    -{\rm Re}\Big[
    e^{-i2\omega t} 
    e^{  i2\omega |\xvec|} 
    \left( -\tilde{\varrho}^{2}_{(1)}(\kvec)
    + \tilde{\bm{j}}^{2}_{(1)}(\kvec)\right)
    \Big] \nonumber
    \Bigg),\\
    &\qquad \textrm{where} \qquad \kvec = \omega \xhat \,,
\end{align}
where  $\tilde{f}(\bm{k})$ is the spatial Fourier transform of $f(\xvec)$. Using the specific forms of the charge and current densities in \eqref{eq:current}, we have \mbox{$\tilde{\varrho}_{(1)}(\bm{k})= -i\g  \omega\tilde{\varphi}(\omega)\xhat\cdot\bar{\bm{B}}$} and
\newline
\mbox{$\tilde{\bm{j}}_{(1)}(\bm{k})   =  -i\omega g_{a\gamma}  \tilde{\varphi}(\omega) \bar{\bm{B}} +ig_{a\gamma}\omega\tilde{\varphi}(\omega) i\xhat\times \bar{\bm{E}}$}, which yields
\begin{align}
\label{eq:dipoleEB}
\frac{dP^\gamma_{(2)}}{d\Omega}  &=
\frac{g_{a\gamma}^2\omega^4\tilde{\varphi}^2(\omega)}{32\pi^2}\Big[
\left(\xhat\times\bar{\bm{B}}\right)^2
+\left(\xhat\times\bar{\bm{E}}\right)^2
-2\xhat\cdot\left(\bar{\bm{E}}\times\bar{\bm{B}}\right)
\Big]
\left(1+\cos\left(2\omega t-2\omega|\xvec|\right)\right)
.
\end{align}
The radiation spectrum is a delta function in frequency, with the radiation emitted at the frequency $\omega$. The spatial pattern of radiation energy density is shown in Fig.~\ref{fig:dipoleRad}. 

\begin{figure}[t]
\begin{center}
\includegraphics[width=0.6\textwidth]{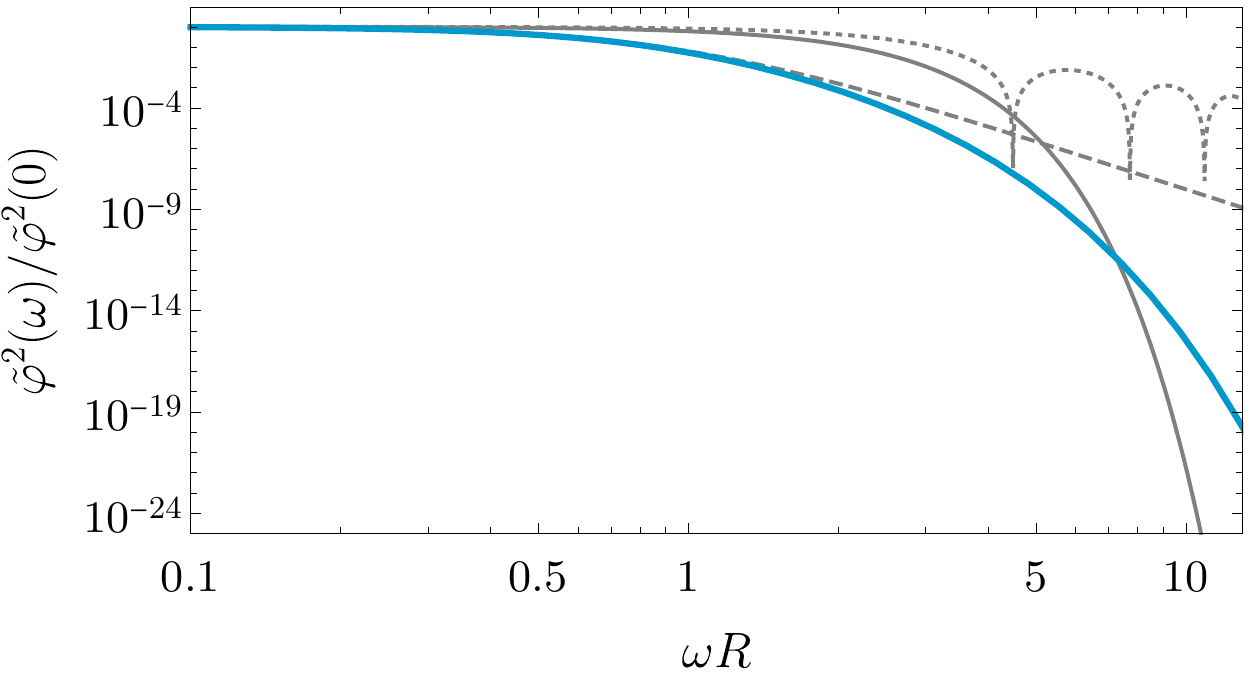}
\end{center}
\caption{The form factor $F(\omega)=\tilde{\varphi}^2(\omega)/\tilde{\varphi}^2(0)$, where $\omega$ is the frequency of oscillation of the axion field (and of the emitted electromagnetic radiation), and $\tilde{\varphi}(\omega)$ is the Fourier transform of the soliton's spatial profile at $k=\omega$. This form factor determines the amplitude of the dipole radiation, and has a very strong dependence on the radius of the soliton.  The blue curve corresponds to the sech profile $\propto \sech(r/R)$, with the correct exponential behaviour at large $R$. The solid gray curve is for a Gaussian profile $\propto e^{-r^2/R^2}$, the dashed one for an exponential profile $\propto e^{-r/R}$ with a cusp at the origin, and the dotted line correspond to a top-hat profile of the soliton with radius $R$. While the form factor is identical at small $\omega R$ for the different profiles, it is very sensitive to the profile choice at large $\omega R$. 
} 
\label{fig:FourierProfiles}
\end{figure}

The total power emitted, and its time average are given by 
\begin{equation}\label{eq:dipole1}
\begin{split}
P^\gamma_{(2)} &= \frac{g_{a\gamma}^2\omega^4\tilde{\varphi}^2(\omega)}{12\pi}
\Big(
\bar{\bm{B}}^2
+\bar{\bm{E}}^2
\Big)
\left(1+\cos\left(2\omega t-2\omega|\xvec|\right)\right)
,\\ 
\langle P^\gamma_{(2)}\rangle_t &= \frac{g_{a\gamma}^2\omega^4\tilde{\varphi}^2(\omega)}{12\pi}
\Big(
\bar{\bm{B}}^2
+\bar{\bm{E}}^2
\Big)
.
\end{split}
\end{equation}
It is important to note that the emitted power is proportional to the squared Fourier transform $\tilde{\varphi}(\omega)$ of the oscillon radial profile evaluated at $\omega$ (which is the frequency of the oscillon, and that of the emitted electromagnetic radiation): 
\begin{align}
\tilde{\varphi}(\omega)= 
\int \! \ud^3\yvec \, \varphi(\yvec)e^{-i\omega\xhat\cdot \yvec } 
=
\frac{4\pi}{\omega}\int_{0}^{\infty} \! \ud r \Big[r\sin(\omega r)\varphi(r)\Big]
\;,
\end{align}
for a spherically-symmetric oscillon.  
Then the ratio $F(\omega) = \tilde{\varphi}^2(\omega) / \tilde{\varphi}^2(0)$ is a form factor for the oscillon profile.  If the wavelength of the radiation is large compared to the scale radius of the oscillon, $R \ll \lambda = \pi / \omega$, then the form factor approaches $F(\omega) \approx 1$ as $\omega \to 0$, corresponding to radiation from a point-like dipole.  Shorter wavelength radiation probes the structure of the oscillon and $F(\omega) \to 0$ as $\omega \to \infty$.  This behavior is illustrated in Fig.~\ref{fig:FourierProfiles} for a few representative oscillon profile functions. 

Using the profile $\varphi(r)=\varphi_0\sech(r/R)$, we get
\Beq
\label{eq:profileF}
\tilde{\varphi}(\omega)
=&\frac{\pi^3\varphi_0 R^2}{\omega}\frac{\tanh(\pi\omega R/2)}{\cosh(\pi\omega R/2)}\approx
\frac{2\pi}{\omega^3}\varphi_0(\pi\omega R)^2 e^{-\pi\omega R/2},
\Eeq
where the second equality assumes $\omega R\gtrsim 2$. When the radius of the axion configuration $R\sim \omega^{-1}$, there is no suppression of the emitted power from $\tilde{\varphi}({\omega})$. However, when $R\gg \omega^{-1}$ we get an exponential suppression.  We have checked that the exponential suppression also exists for numerically obtained spatial profiles for dilute axion stars where $\omega R\gg 1$. 

The physical origin of this suppression is destructive interference between the emitted electromagnetic waves which are emitted in phase from different locations within the oscillon. Also see discussion of coherence and interference in section \ref{sub:coherence}. Note that this suppression is more severe than the suggested by \cite{Bai:2017feq}, where a power law suppression is obtained because of to a cusp in their $\varphi(r)$ at the origin. This can make a rather large difference in the radiated power even for $\omega R\gtrsim \rm{few}$. Compare the blue curve for the sech profile with the dashed gray curve for an exponential profile with a cusp at the origin.\footnote{Note that one can define the scale $R$ for different profiles (approximately) in terms of the radius $R_{90}$ which encloses $90\%$ of the soliton mass. For the exponential and sech profiles we find $R\approx 0.4R_{\rm 90}$, whereas for a Gaussian profile $R\approx 0.8 R_{\rm 90}$ when $R$ is sufficiently large. } Also see Sec.~\ref{sec:axion_stars} for further discussion of the expected form of the axion star profiles.
\subsubsection*{Summary of dipole radiation}
Finally, to make the dipole nature of the radiation apparent, let us set the background electric field to zero. In this case
\begin{align}
\label{eq:dipoleB}
\left\langle\frac{dP^\gamma_{(2)}}{d\Omega}\right\rangle_t =
\frac{g_{a\gamma}^2\omega^4\tilde{\varphi}^2(\omega)}{32\pi^2}\bar{\bm{B}}^2\sin^2\theta\approx\frac{(g_{a\gamma}\varphi_0)^2}{8\omega^2}(\pi \omega R)^4e^{-\pi\omega R}\bar{\bm{B}}^2\sin^2\theta,
\end{align}
where $\theta$ is the angle with respect to the $\bar{\bm{B}}$ direction. The same formula holds for the electric field also. The second equality is a good approximation for $\omega R\gtrsim 2$ for the $\sech$ profile. To get significant emitted power, it is essential to have $R\omega$ not be too large, and $\g\varphi_0$ not be too small, which provides motivation for considering dense axion stars and oscillons. At the same time, it is also beneficial to have a small $\omega\sim m_a$ which pushes us towards pursuing lighter axions.

\subsubsection{Higher orders in $\g\varphi_0$: beyond dipole radiation}
\label{sec:higherorder}
Our organization of the calculation using powers of $\g\varphi_0$ is fraught with subtleties as we go beyond the leading order in $\g\varphi_0$, with the system best dealt with non-perturbatively  using Floquet theory (with a large number of coupled degrees of freedom). However, to appreciate these subtleties, we try to follow our nose and proceed with the calculation order by order in $\g\varphi_0$. While we will be unable to complete the calculation, the set up also provides some physical insight into how  the radiated power deviates from the dipole estimate of the previous section as we increase the coupling strength.

The field equations, charge and current densities are given by
\begin{align}
\label{eq:EBJn}
&\ddot {\bm{E}}_{(n)} - {\bm{\nabla}}^2 {\bm{E}}_{(n)} = - {\bm{\nabla}}\rho_{(n)} -\dot {\bm{J}}_{(n)}\,,&\qquad&\ddot {\bm{B}}_{(n)} - {\bm{\nabla}}^2 {\bm{B}}_{(n)} =  {\bm{\nabla}}\times {\bm{J}}_{(n)}\,,\\
&\rho_{(n)}=-\g \nabla\phi\cdot \bm{B}_{(n-1)},&\qquad &\bm{J}_{(n)}=\g \left(\dot{\phi}\bm{B}_{(n-1)}+\nabla\phi\times\bm{E}_{(n-1)}\right)\,.
\end{align}
for $n\ge 1$. Recall that $(n)$ denotes the order in $\g\varphi_0$, ${\bm{E}}_{(0)}=\bar{\bm{E}}$ and ${\bm{B}}_{(0)}=\bar{\bm{B}}$ are assumed to be constants, and $\phi=\varphi(r)\cos\omega t$. We will continue ignoring backreaction of the produced electromagnetic fields on the axion field configuration in this subsection.\footnote{Note that the change in the axion field configuration due to backreaction $\delta\phi/\varphi_0\sim \g^2\bar{B}^2/m^2\ll 1$.} Order by order this represents a system of periodically forced oscillators. We can use these to understand the possible frequency structure of the fields at different orders (ignoring resonances for the moment). Since $\phi$ oscillates with a frequency $\omega$, so do $(\rho_{(1)},\bm{J}_{(1)})$, which in turn source $(\bm{E}_{(1)},\bm{B}_{(1)})$ which also oscillate with a frequency $\omega$. However, because of the products of oscillating terms coming from $\phi$ and oscillating electromagnetic fields, $(\rho_{(2)},\bm{J}_{(2)})$ will include frequency components $0\omega$ and $2\omega$. Similarly, $(\rho_{(3)},\bm{J}_{(3)})$ will contain $\omega$ and $3\omega$ and so on. 

The above arguments reveal that if we are interested in the radiated electromagnetic fields at $\mathcal{O}[(\g\varphi_0)^n]$, they will contain multiple frequencies. Conversely, if we want to consider fields with a fixed frequency, they will contain terms with many different orders in $\g\varphi_0$. This latter fact does mean that there is a possibility that the generated electromagnetic fields (and power) at a given frequency, or in total, can be enhanced or decreased as we go to higher couplings. That is, the power radiated can be non-monotonic in the coupling (when the coupling is not too small), and its dominant frequency content might also change with coupling strength. We observe these effects in our numerical simulations. Finally, note that fields with higher frequencies beyond $\omega$ always come with higher powers in $\g\varphi_0$; this is because higher frequencies are sourced by the electromagnetic fields already sourced by the axion field configuration. Again, we confirm this behavior in the simulations.

The above discussion is incomplete because we ignored the possibility of resonances that should be present in a system with periodic forcing terms. These resonances make it notoriously difficult to carry out our perturbative scheme for long time scales.
We can get a rough idea of the difficulties and subtleties by trying to solve the \eqref{eq:EBJn} equations in momentum space.
First, let us consider the $n=1$ case: 
\begin{align}
&\ddot{\bm{B}}_{(1)}(t,\pvec)+p^2{\bm{B}}_{(1)}(t,\pvec)= -i\omega g_{a\gamma}  \tilde{\varphi}(p)\pvec\times  \bar{\bm{B}}\sin(\omega t)
,
\end{align}
with ${\bm{B}}_{(1)}=0$ and $\dot{\bm{B}}_{(1)}=0$ at $t=0$.
The general solution is
\begin{align}
{\bm{B}}_{(1)}(t,\pvec)=\left\{
\begin{array}{ll}
\vspace{1.5ex}
-i\omega g_{a\gamma}  \tilde{\varphi}(p)\pvec\times  \bar{\bm{B}} \frac{p\sin(\omega t)-\omega\sin(pt)}{p(p^2-\omega^2)}, & p\neq \omega \\
-i\omega g_{a\gamma}  \tilde{\varphi}(p)\pvec\times  \bar{\bm{B}} \frac{\sin(\omega t)-t\omega\cos(\omega t)}{2\omega^2}, & p= \omega
\end{array}
\right.
.
\label{eq:Bp}
\end{align}
As we see, the result is periodic and bounded, except when $p$ equals $\omega$. This is just the behavior of a periodically forced harmonic oscillator.
As expected, for $p=\omega$, there is a term that is linear in $t$ (secular growth). After sufficient time, such a term will dwarf the zero order terms.
This in turn can limit the reliability of the perturbative expansion we used in the first place. To maintain the validity of the perturbative expansion, besides requiring a small $\g\varphi_0$, one should further restrict ourselves to small times. This scenario with secular terms is reminiscent of the challenge of solving the Mathieu equation via perturbative methods, and in principle, there exist mathematical tools to deal with such situations. For example, one can go beyond the naive perturbation theory (\ref{eq:pertur}), and resort to the Renormalization Group \cite{Chen:1995ena} or resurgent resummation \cite{Dunne:2016qix}. But the question here is more complicated than in the Mathieu equation because of the vast number of coupled momentum degrees of freedom.

Furthermore, there is another subtlety. While the individual mode for $\bm{B}_{(1)}(p=\omega)$ has a secular term, when we obtain the fields in position space via a Fourier transform, the secular term disappears. Note that the secular term above can be reached by the expression of $p\neq \omega$, in the limit $p\to \omega$, and therefore this imposes no pole or singular point in the momentum integral. 

Nevertheless, there are good reasons to believe the secular terms will appear at high order in $\g\varphi_0$ terms for the fields. One reason is that Floquet theory predicts (and we observe in simulations) the existence of exponentially growing solutions that can be constructed out of solutions with different power law (unbounded) in time dependencies at various orders. More generally, such terms can combine in non-trivial ways to give real and imaginary Floquet exponents corresponding to bounded and unbounded solutions.

\section{Results of numerical lattice simulation}
\label{sec:lattice}

The axion-photon system can be simulated numerically \cite{Adshead:2015pva,Adshead:2019lbr,Figueroa:2017qmv,Cuissa:2018oiw,Amin:2020vja}.
In general, when there exist charged matter fields, the usage of the electromagnetic potential $A_{\mu}$ is unavoidable, as the gauge covariant derivatives require $A_{\mu}$ explicitly.
But in the simple axion-photon system where there is no charged field, the electromagnetic scalar potential $\phi$ and vector potential $\bm{A}$ are not necessary.
Instead, we can directly evolve the electric field $\bm{E}$ and the magnetic field $\bm{B}$ in the axion background through the Maxwell's equations, and as a byproduct, there is no gauge fixing needed. We will present the details of our numerical scheme in a separate paper.
\\ \\
\noindent{\bf Simulation Parameters and Initial Conditions}: Our benchmark simulation has a physical volume $m^3V=64\times 64\times 64$, with $N^3=160^3$ lattice sites, and the resolution is $m\ud x=0.4$. We also used $N^3=320^3$ for convergence tests (for more details, see \cite{Amin:2020vja}). We typically run our simulations up to $mt_{\rm max}=50$. We also used $mt_{\rm max}=100$ when using an eight times larger simulation volume. We employ periodic boundary conditions but make sure that we do not have our results contaminated by radiation cycling through the box.

For initial conditions we start with the axion field in a solitonic configuration of the form $\phi(t)=\varphi_0 \sech(r/R)\cos(\omega t+\theta_0)$ with $\theta_0=-\pi/2$. That is, $\phi(t=0)=0$  and $\dot{\phi}(t=0)= \omega\varphi_0 \sech(r/R)$. We include either a constant magnetic field or a constant electric field through the box. For our fiducial values of our parameters, we use 
\Beq
\varphi_0= 2.6 f\, , \quad \omega = 0.82 m \, , \quad R= 1.6m^{-1} \, , \quad \textrm{and} \quad \bar{E}=10^2m^2 \ \textrm{or} \ \bar{B}=10^2m^2\,.
\Eeq
With these above values (see Eq.~\eqref{eq:EffCoup_def}):
\Beq
\mathcal{C}=\g\varphi_0 \omega R/4\approx 0.85(f\g)\,.
\Eeq
The soliton parameters are  consistent with those of dense solitons  found in \cite{Zhang:2020bec}, although the precise values can differ based on the functional form used to fit the true profile. Apart from transients, our results are  insensitive to the chosen values of $\bar{E}$ and $\bar{B}$ apart from a trivial scaling of the radiated power in the $\Cc\ll1$ regime, and a change in logarithmic time scale of backreaction (see below) in the $\Cc\gtrsim 1$ regime.\footnote{Note that we do not need to add seed fluctuations in the electromagnetic field (unlike the case in \cite{Amin:2020vja}), since the background electromagnetic field in presence of the axion-configurations sources the electromagnetic field fluctuations. } The chosen values of the background fields are for numerical convenience. We also varied these parameters within factors of two or even an order of magnitude to delineate general statements from those that are sensitive to this particular choice of fiducial parameters.

While we focus on dense solitons, we could have carried out simulations in the dilute soliton regime as well. However, the exponential suppression expected from \eqref{eq:dipole1} would make this uninteresting (at least for dipole radiation, though not necessarily for the case with parametric resonance \cite{Hertzberg:2018zte}). 
\\ \\
\noindent{\bf Backreaction Considerations}: For most of our simulations, it is unnecessary to evolve the axion field using its equation of motion numerically (although it is still necessary to solve for the electromagnetic fields numerically for $\Cc\sim 1$). That is, the soliton sources electromagnetic fields, but it is not significantly affected by them. To see this, recall that the energy extracted from a dense soliton with $\Cc\ll 1$ grows linearly with time  with $P^\gamma\sim 10\bar{B}^2(f\g)^2/m^2$ (see \eqref{eq:dipole1}). Hence it will take
\Beq\label{eq:tbr}
mt_{\rm br}\sim \frac{mM_{\rm sol}}{P^{\gamma}}\sim 10\frac{m^2}{\g^2\bar{B}^2}\gg mt_{\rm max}\,,\qquad \Cc\ll 1\,,
\Eeq
for backreaction on the soliton to be relevant. We have used $M_{\rm sol}\sim 10^2f^2/m$ above for the energy of a dense soliton. Note that since even for the strongest fields possible around neutron stars $\bar{B}\lesssim m_e^2\sim 10^{-1}{\rm MeV}^2$, and with $\g\lesssim10^{-10}\rm GeV^{-1}$ and $m\gtrsim 10^{-7}{\rm eV}$, we have $m/(\g \bar{B})\gg 1$. Hence, we can safely ignore backreaction on the axion field configuration in our simulations when $\Cc\ll 1$ and $mt_{\rm max}\lesssim 10^2$. 

Note that another way of thinking about backreaction is at the level of the equation of motion for the axion field (first equation in \eqref{eq:KGMax}). The correction to the axion field evolution due to the source term $\g \bm{E}\cdot\bm{B}$ is given $\delta\phi/\varphi_0\sim \g^2\bar{B}^2/m^2\ll 1$. This is essentially the same ratio that appears in the discussion above.
\begin{figure}[t]
\begin{center}
\begin{minipage}[c]{0.6\textwidth}
\includegraphics[width=1\textwidth,trim=0.4cm 0.8cm 0.4cm 2cm, clip]{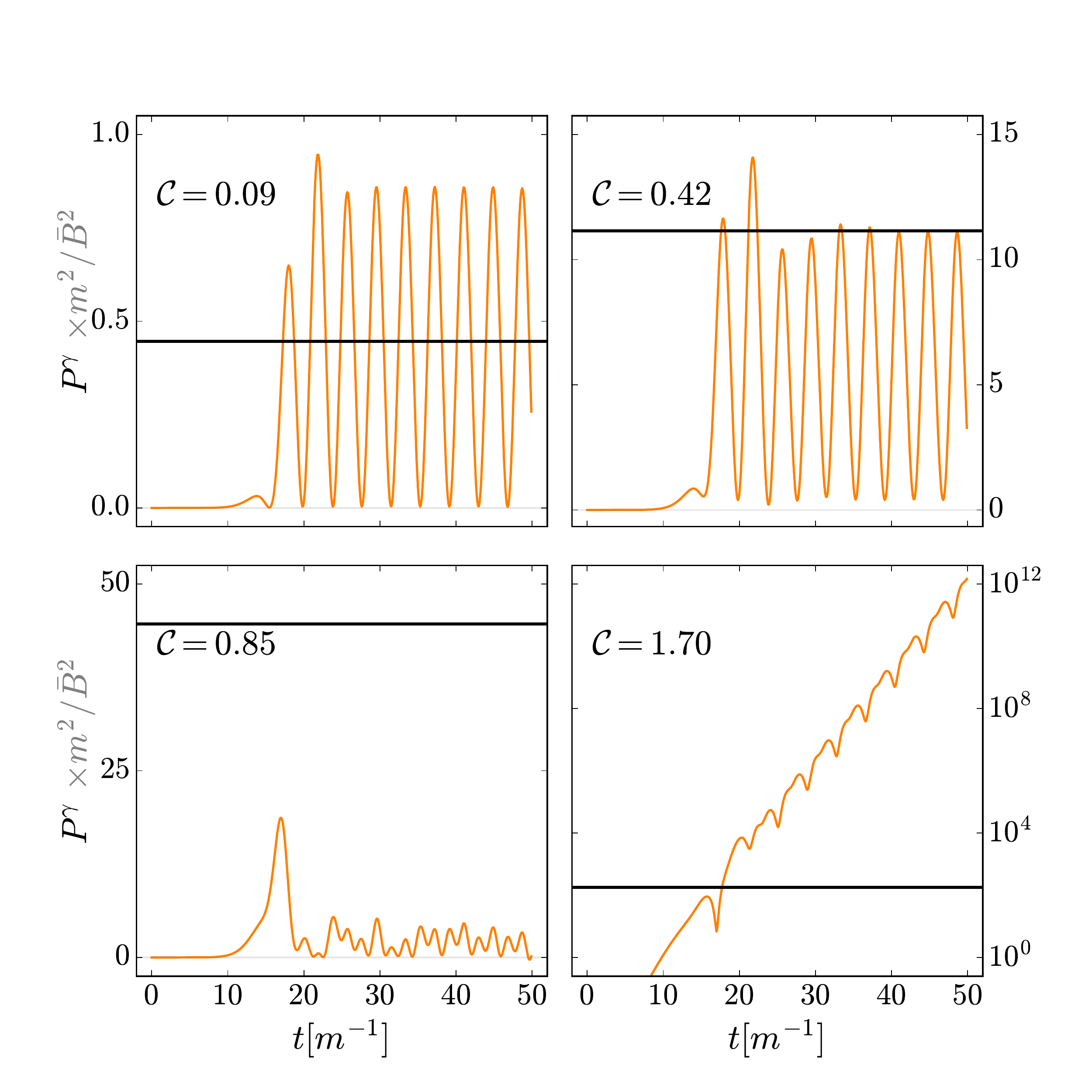}
\end{minipage}
\begin{minipage}[c]{0.38\textwidth}
\includegraphics[width=1\textwidth,trim=1cm 0cm 0cm 0cm, clip]{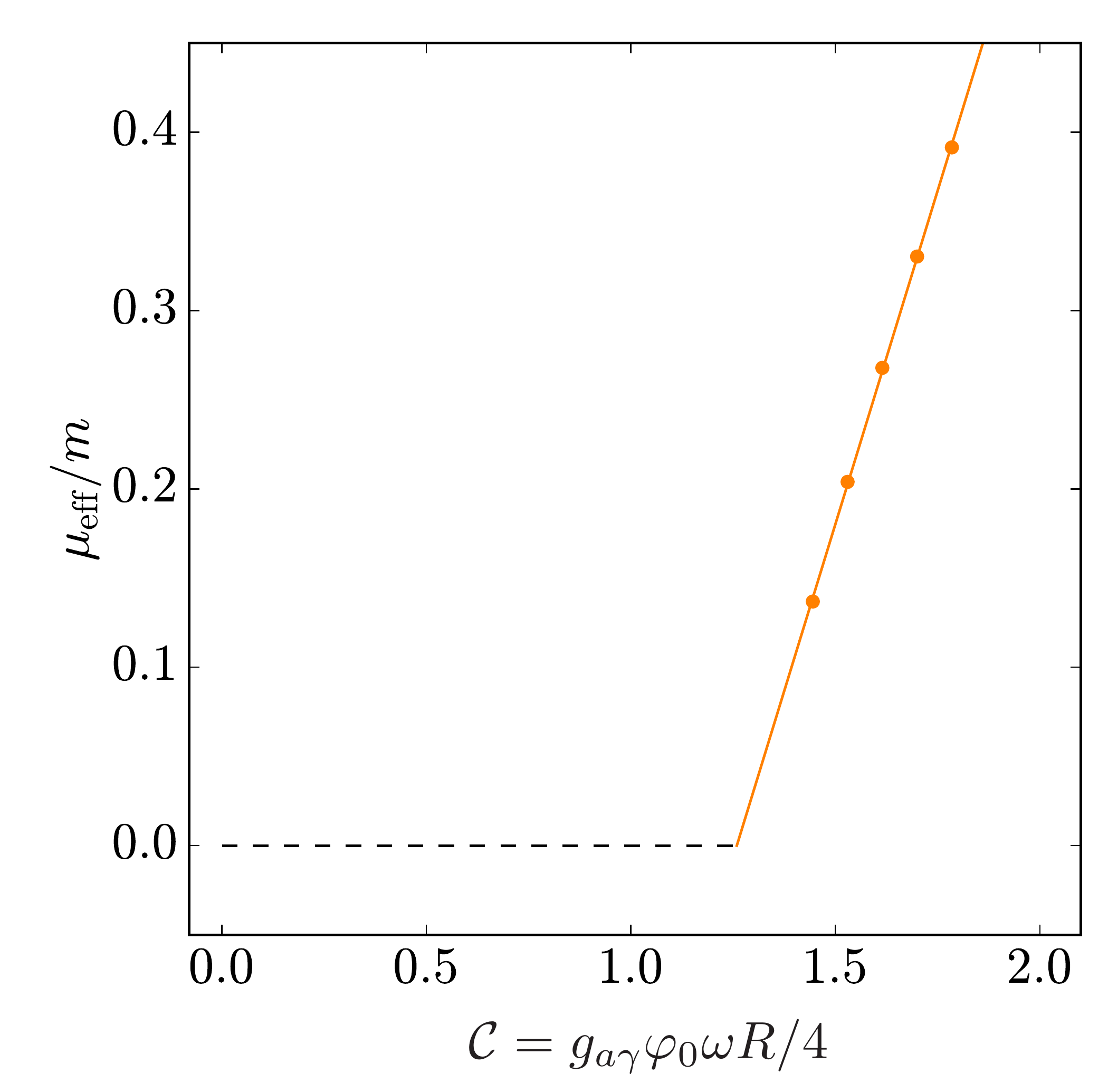}
\end{minipage}
\end{center}
\caption{
(Left) The radiated power versus time for the different effective dimensionless coupling $\Cc=\g\varphi_0\omega R/4=0.09,~0.42,~0.85$ and $1.7$ due to a dense soliton in a constant magnetic field background. Notice on the last plot (lower right) the logarithmic vertical scale and the unbounded solution, while the other three are bounded.
(Right) The exponent $\mu_{\rm eff}$ extracted from the exponential growth of the radiated power.
There exists a linear relationship between $\mu_{\rm eff}$ and $\Cc$ in the unbounded region.
The $\Cc_{\rm critical}\approx 1.3$ can be read off as the zero point of the linear relationship, with $\mu_{\rm eff}/m\approx 0.75\Cc$. We caution that the precise numerical coefficients depend on the details of the soliton configuration. For the above plots we are dealing with a dense soliton $\varphi_0\sim f$ and $R\sim {\rm few}\times m^{-1}$ and $\omega\lesssim m$.
}
\label{fig:phases}
\end{figure}

For $\Cc\gtrsim 1$, the exponential growth in the radiated power can lead to backreaction on the soliton within $mt_{\rm max}$. At the end of this section, we provide simulation results where backreaction eventually shuts down the resonant electromagnetic field production.
\\ \\
\noindent{\bf Numerically Calculated Power}: The main output from our simulations will be the radiated power in electromagnetic fields. We define this radiated power as the surface integral of the Poynting vector over a spherical surface whose radius is much larger than the size of our soliton. In our set-up, we compute the luminosity by a sum
\begin{align}
P^\gamma \equiv |\xvec|^2\int \! \ud \Omega \,  \xhat\cdot(\bm{E}\times \bm{B})
= \frac{4\pi |\bm{r}|}{{\mathcal N}}\sum_{j=1}^{\mathcal N}\bm{r}_j\cdot\left[\bm{E}_j\times \bm{B}_j\right]
\end{align}
where the sum is over all sites of index $j$ with a distance within $(r-\epsilon,r+\epsilon)$, given $\epsilon\ll m^{-1}$. Note that we exclude $\bar{\bm{E}}$ or $\bar{\bm{B}}$ when we compute the Poynting vector since these background fields are not part of the radiation that escapes to infinity. 
In our simulations, the radiated power is measured at $mr=16$ with $m\epsilon=0.1$. Note that for $mt_{\rm max}=50$, the radiation does not have sufficient time to cycle through our periodic box and contaminate the radiated power calculation.

In general, we found that the numerically calculated power was not very sensitive to the lattice size or the radius of the sphere where we calculated the radiated power as long as this radius $\gg R$. Our finite lattice spacing, $m\ud x$, leads to a slightly smaller numerically evaluated power in comparison to the power calculated in the continuous limit (when such a calculation is possible). For $m\ud x=0.4$, the discrepancy with the analytic expectation is $\sim 1\%$ for $\Cc\lesssim 1$.

\begin{figure}[t]
\begin{center}
\includegraphics[width=\textwidth]{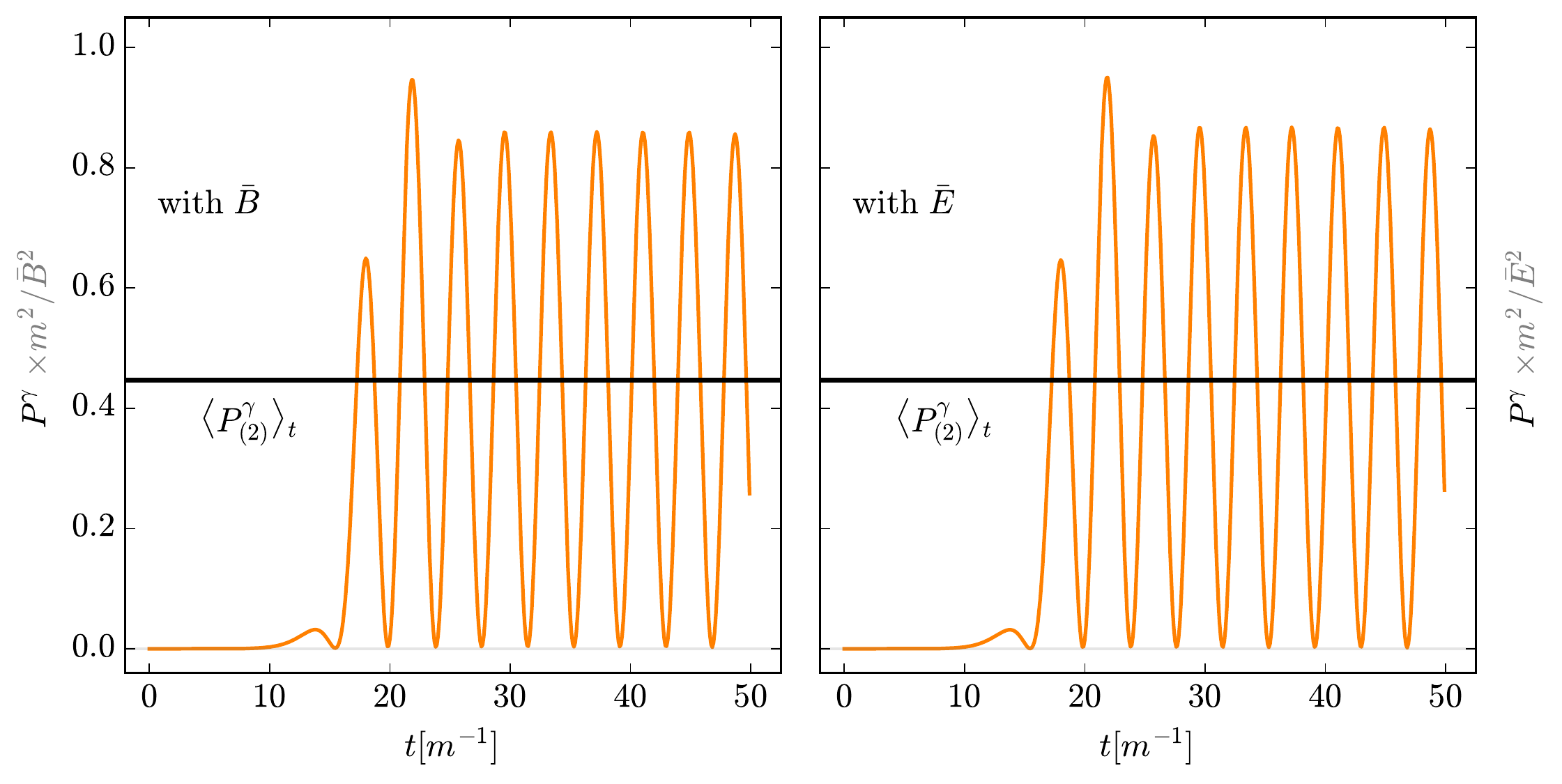}
\end{center}
\caption{
The radiated electromagnetic power from a dense soliton in constant background $E$ or $B$ field as function of time for $\Cc=0.09$. In this regime the radiation is expected to be described well by dipole radiation $P_{(2)}^\gamma$ provided in \eqref{eq:dipole1}. The black lines are the analytic results for the time-averaged power which matches nicely with the numerical results. Note the frequency of $2\omega$ is expected from analytics as well.}
\label{fig:smallG}
\end{figure}

\subsection{Bounded vs. unbounded radiating fields}
As we discussed at the beginning  of Section \ref{sec:Analytic}, we expect periodic solutions for $\Cc\ll 1$ and exponential growing ones for $\Cc\gtrsim 1$ based on Floquet theory. To confirm this behaviour, we numerically solve for $\bm{E}$ and $\bm{B}$ fields sourced by the same soliton configuration (with our fiducial values of $R$, $\varphi_0$ and $\omega$), but different $f\g$. 

In Fig.~\ref{fig:phases} (left panel) we plot the power radiated as a function of time for different $\g$. Note that the power radiated is constant for $\Cc\lesssim 1$ but increases exponentially as $P^\gamma\propto e^{2\mu_{\rm eff} t}$ for $\Cc\gtrsim 1$ as expected. Also note the different scales on the vertical axes for different parts of the panel. In the right panel of  Fig.~\ref{fig:phases}, we plot the maximum Floquet exponent $\mu_{\rm eff}$ from the numerically obtained time dependence of the radiated power. This plot reveals that 
\Beq
\mu_{\rm eff}/m\approx 0.75\times \Cc\,\qquad\textrm{for}\qquad \Cc\ge \Cc_{\rm crit}\approx1.3\,.
\Eeq
Note that the Floquet exponent is a property of the axion configuration and coupling $\g$ (though the combination in $\Cc$), but is independent of the presence or absence of background electromagnetic fields. While we expect $\Cc_{\rm crit}\sim 1$, its precise value and the numerical coefficient appearing in the expression for $\mu_{\rm eff}$ will depend on the details of the soliton solution.

\begin{figure}[t]
\begin{center}
\includegraphics[width=0.75\textwidth]{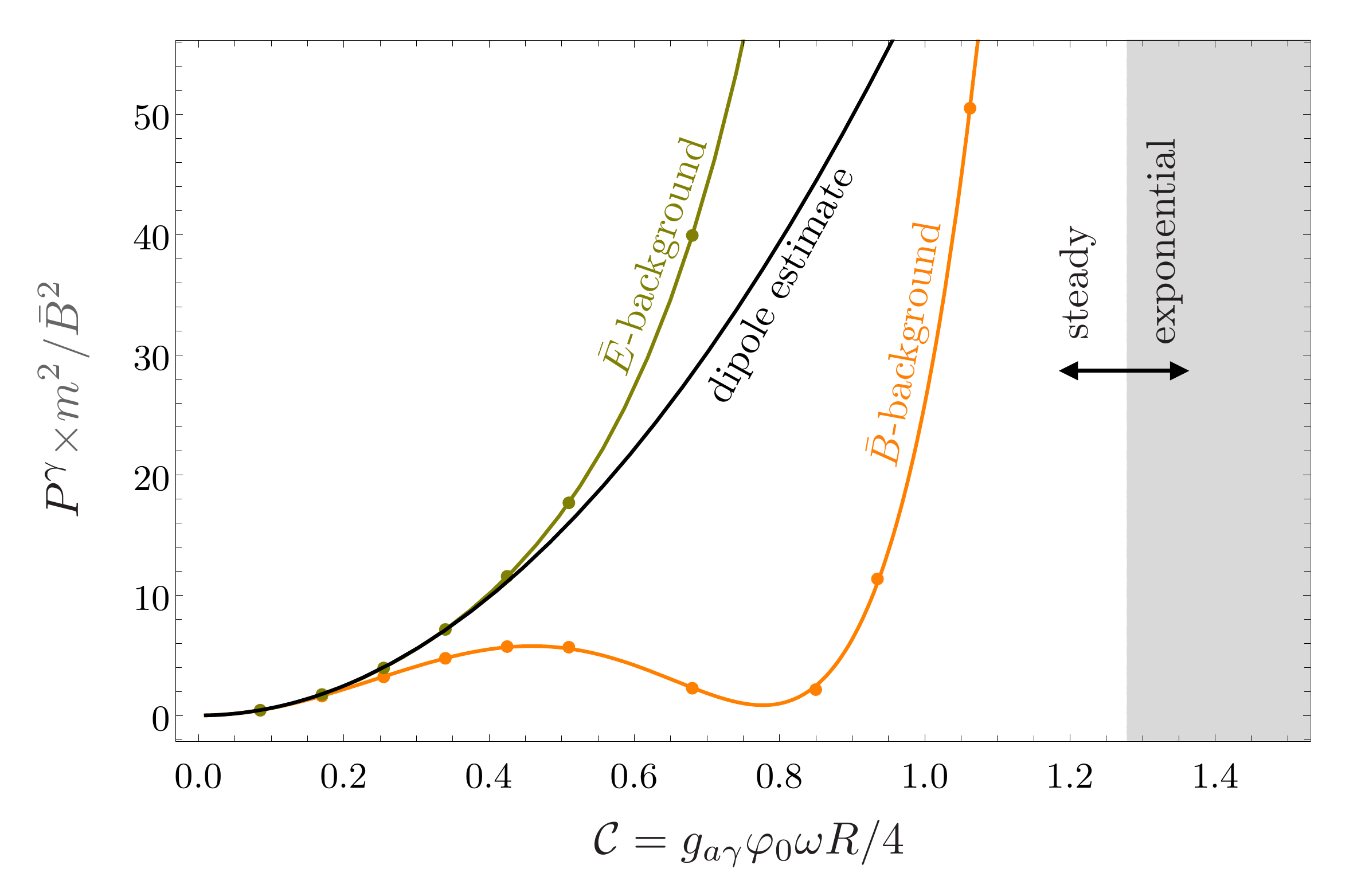}
\end{center}
\caption{The dependence of the time-averaged radiated power $\langle P^\gamma\rangle_t$ on the effective dimensionless coupling $\Cc=\g \varphi_0 \omega R/4$. The above plot is based on a dense soliton of field amplitude $\varphi_0=2.6 f$, frequency $\omega=0.82 m$ and radius $R=1.6m^{-1}$. For changing $\Cc\approx 0.85(f\g)$, only $f\g$ is varied. The black line is the dipole estimate $\langle P_{(2)}^\gamma\rangle_t$ from eq.~\eqref{eq:dipole1}. The orange and green dots show the numerically evaluated $\langle P^\gamma\rangle_t$ for $E$ and $B$ field backgrounds respectively. Note that the numerics agree with the dipole estimate at $\Cc\ll1$ as expected. The deviation becomes more and more pronounced as we move from $\Cc\ll 1$ towards $\Cc\sim 1$. In particular, note the difference in the radiated power between $E$ and $B$ field backgrounds.  For the $B$ background, note the significant suppression of the radiated power compared to the dipole estimate and the non-monotonic behaviour with $\Cc$. Finally, as $\Cc>\Cc_{\rm crit}=1.3$ (grey shaded), we have an exponentially growing (in time) power due to parametric resonance.}
\label{fig:lum_g}
\end{figure}

\subsection{Small coupling: dipole estimate}

In Section \ref{sec:dipole}, we provided an analytic calculation for the power radiated by the soliton configuration in the presence of background $E$ and/or $B$ fields. This result is expected to hold for $\Cc\ll 1$.

We confirm this expectation in detail with numerical simulation for $\Cc\approx0.09$. In Fig.~\ref{fig:smallG}, we present the time-dependent power radiated by the soliton in the presence of a constant background $B$ field (left) and $E$ field (right). The radiated power oscillates with a frequency $2\omega$, consistent with our analytic result in the first line of \eqref{eq:dipole1}. Moreover, the magnitude of the time averaged power is also consistent with our analytic calculation in the second line of \eqref{eq:dipole1} (horizontal black lines). Finally, the spatial pattern of the radiated energy density is consistent with dipole radiation as predicted in eq.~\eqref{eq:dipoleB} (see Fig.~\ref{fig:dipoleRad}). 

While we do not do so here, we can easily include both electric and magnetic field backgrounds together to confirm eq.~\eqref{eq:dipoleEB}.

\begin{figure}[p!]
\begin{tabular}{c}
\hspace{-0.5cm} \vspace{-0cm}
\includegraphics[width=1\textwidth,trim=0.1cm 0.2cm 0.1cm 0cm, clip]{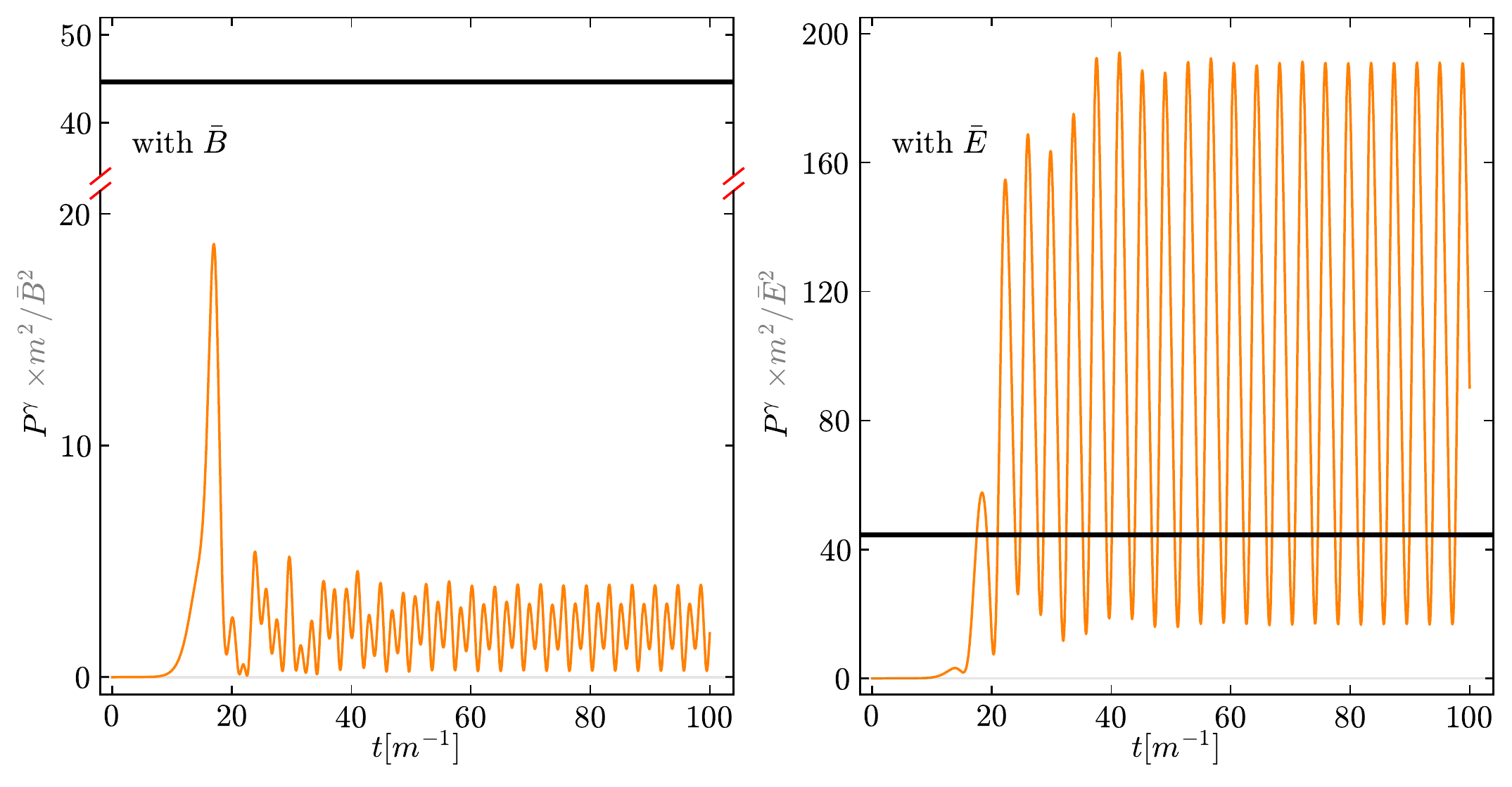}
\end{tabular}
\caption{
Power radiated our fiducial dense soliton in a background magnetic (left panel) and electric (right panel) field, for an intermediate  value of the effective dimensionless coupling $\Cc=0.85$. The horizontal black line refers to the value from the dipole estimate, which is the same on both left- and right-hand plots. 
Note that there is a significant suppression of the radiated power in a magnetic field background compared to the dipole estimate, with the radiated power becoming dominated by the $4\omega$ (instead of the $2\omega$) radiation. However, for the electric field background, there is an enhancement compared to the dipole estimate.
}
\label{fig:largeG}
\end{figure}

\begin{figure}[p!]
\begin{tabular}{lr}
\hspace{-0.8cm} \vspace{-0cm}
\includegraphics[width=0.53\textwidth,trim=2.4cm 0cm 3.0cm 0.7cm, clip]{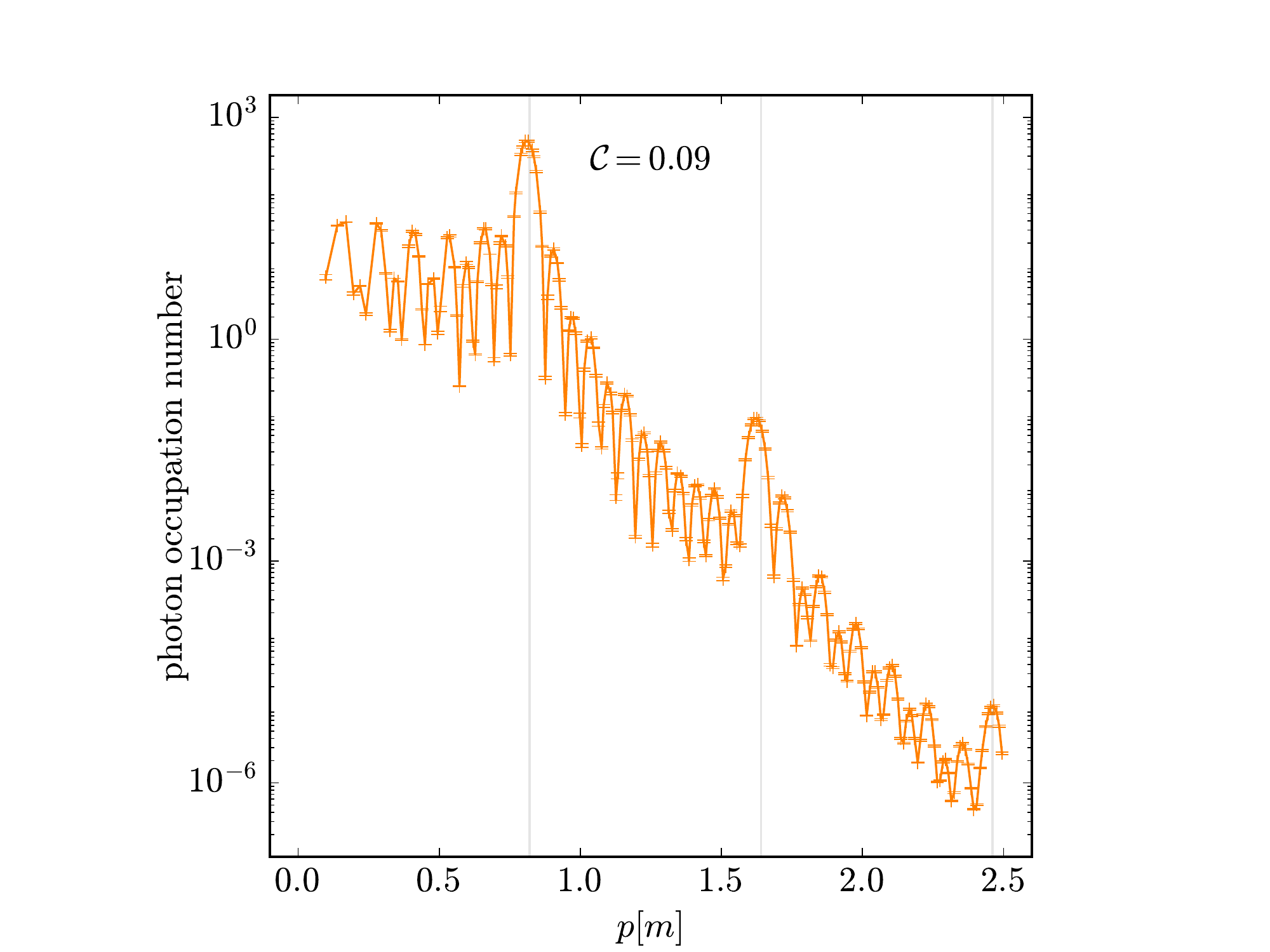} & \hspace{-0.8cm}
\includegraphics[width=0.53\textwidth,trim=2.4cm 0cm 3.0cm 0.7cm, clip]{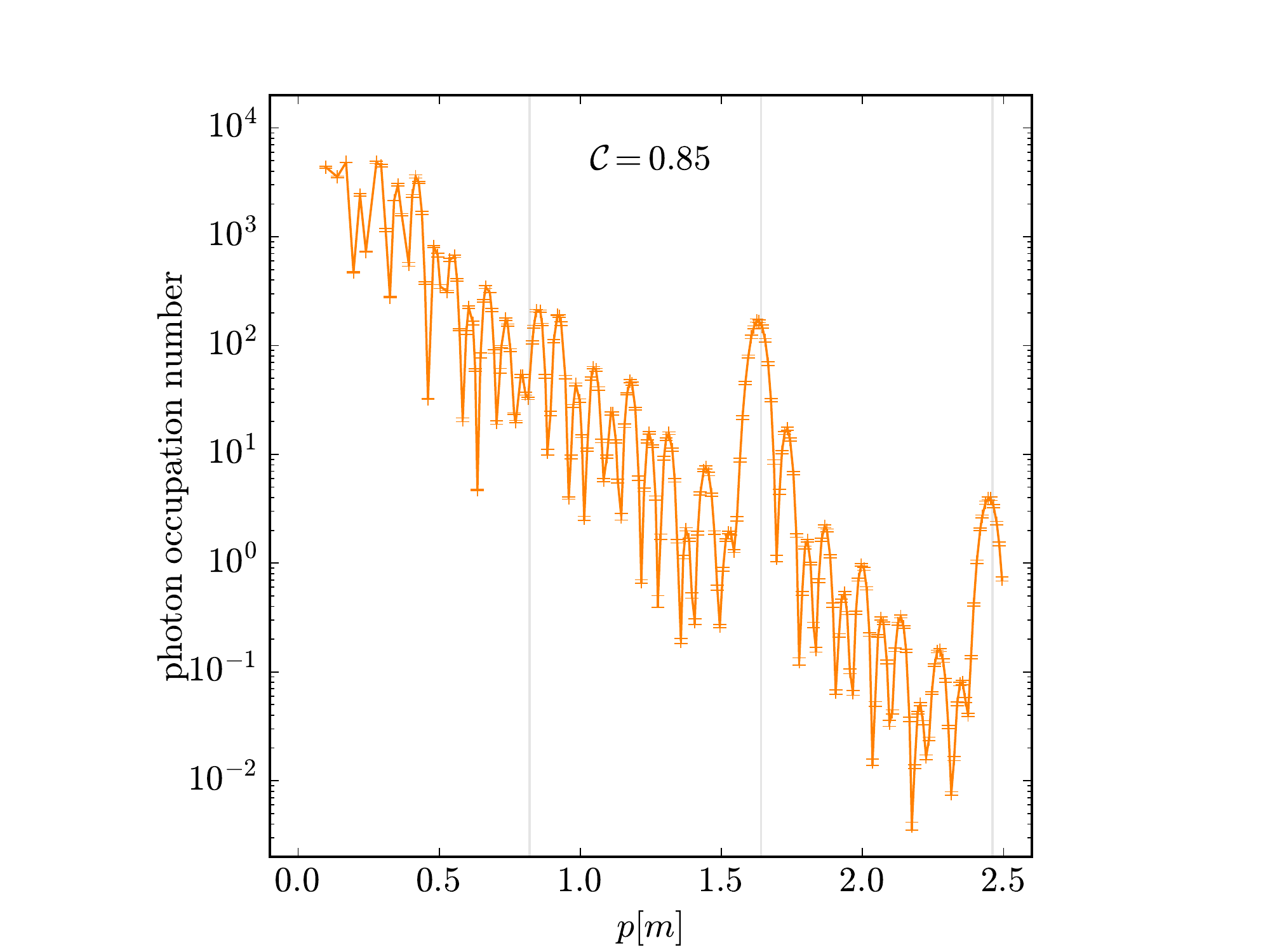}
\end{tabular}
\caption{
The photon particle number distribution in momentum space obtained by taking a Fourier transform of the electromagnetic fields in our simulation volume at a fixed time (for more details, see \cite{Amin:2020vja}).
Three gray lines represent $p=\omega$, $2\omega$ and $3\omega$. Left panel is for $\Cc=0.09$ whereas the right panel is for $\Cc=0.85$. Note that at $\Cc=0.85$, the dominant radiation frequency in $\omega$ is absent, and is connected to the suppression of power in the magnetic field background at this coupling.
}
\label{fig:pNum}
\end{figure}

\subsection{Intermediate couplings}
We now consider the power radiated when $0.1\lesssim \mathcal{C}\lesssim 1$. As we saw in Section \ref{sec:higherorder}, it is difficult to extend the perturbative calculation to include terms that are higher order in $\g\varphi_0$. Numerically, we of course have no issues probing this regime. In this regime, new phenomenon emerge, which have not been reported before to the best of our knowledge.

As we move away from the $\Cc\ll 1$ regime towards $\Cc=\Cc_{\rm crit}$, the radiated power is still constant in time (ie. we have periodic solutions for the radiated fields at each point in space). However, we find that the dipole estimate $\langle P^\gamma_{(2)}\rangle_t$ from eq.~\eqref{eq:dipole1} starts deviating significantly from our numerical results. See Fig.~\ref{fig:lum_g}, where the black curve is $\langle P^\gamma_{(2)}\rangle_t$ whereas the orange and green curves represent the numerically obtained $\langle P^\gamma\rangle_t$ for the case of background $B$ and $E$ fields respectively. In the following, when we vary $\Cc$, we hold all parameters fixed apart from $f\g$.
\\ \\
\noindent{\bf Background B}: For the case of the background $B$ field, we find that the $\langle P^\gamma_{(2)}\rangle_t$ overestimates the power in this regime (see Fig.~\ref{fig:lum_g}). The frequency content of the radiated power continues to be dominated by $2\omega$. As discussed in Section \ref{sec:higherorder}, we believe that this is due to higher order contributions of $\mathcal{O}((\g\varphi_0)^4)$ to the radiated power at frequency $2\omega$. These next-to-leading order in $\g\varphi_0$ contributions have an opposite sign compared to the leading order result. This is confirmed by our fits to $\langle P^\gamma \rangle_t$ as a function of $\Cc(\propto \g$).  As $\Cc\propto \g$ increases further, $\langle P^\gamma\rangle_t$ even shows a non-monotonic behavior -- first increasing with $\g^2$ at small $\g$ and then turning over and decreasing as $\g$ increases further. Increasingly higher order terms in $\g\varphi_0$ can no-longer be ignored as $\Cc\sim 1/2$. 

For our fiducial parameters for the profile, we see that the dipole estimate can differ by more than an order of magnitude as $\Cc$ becomes order unity (see Fig.~\ref{fig:largeG}, left panel). We also find that for certain $\Cc$, even the dominant frequency content of the radiated power can change from $2\omega$ to $4\omega$ signalling a cancellation between various higher order terms in $\g$ at the frequency $2\omega$! In Fig.~\ref{fig:pNum} we show a comparison between the frequency(=wavenumber) content of the radiation in our simulation box at for $\Cc=0.09$ and $\Cc=0.85$. Notice the vanishing of the dominant $k\approx\omega$ as we go from $\Cc=0.09$ to $\Cc=0.85$. 
\\ \\
\noindent{\bf Background E}: For the case of the background $E$ field, we find that the $\langle P^\gamma_{(2)}\rangle_t$ underestimates the power as we move to larger $\Cc$. See Fig.~\ref{fig:lum_g}. The frequency content of the radiated power continues to be dominated by $2\omega$. The detailed reason for difference in behaviour of the radiated power between the background $B$ and $E$ fields is not entirely clear to us. However, we do note that it might be sensitive to the fiducial parameters chosen. By reducing the radius of the soliton, we were able to also find a regime where the $\langle P^\gamma_{(2)}\rangle_t$ overshoots the numerical results even with an $E$ field background.

\subsection{Large coupling with backreaction}
When the coupling $\Cc>\Cc_{\rm crit}\approx 1.3$, we transition to exponentially growing $E$ and $B$ fields. This exponential growth can be rapid enough so that an order unity fraction of the energy of the soliton is extracted from the soliton within the duration of our simulations. The duration to backreaction depends on the initial energy in the soliton, as well as initial conditions. For the present case the background $B$ and $E$ fields generate fluctuations in the electromagnetic fields, which are then enhanced via parametric resonance. 

\begin{figure}[t]
\begin{center}
\begin{tabular}{lr}
\hspace{-1cm}
\includegraphics[width=0.55\textwidth,trim=2cm 0cm 2.5cm 1cm, clip]{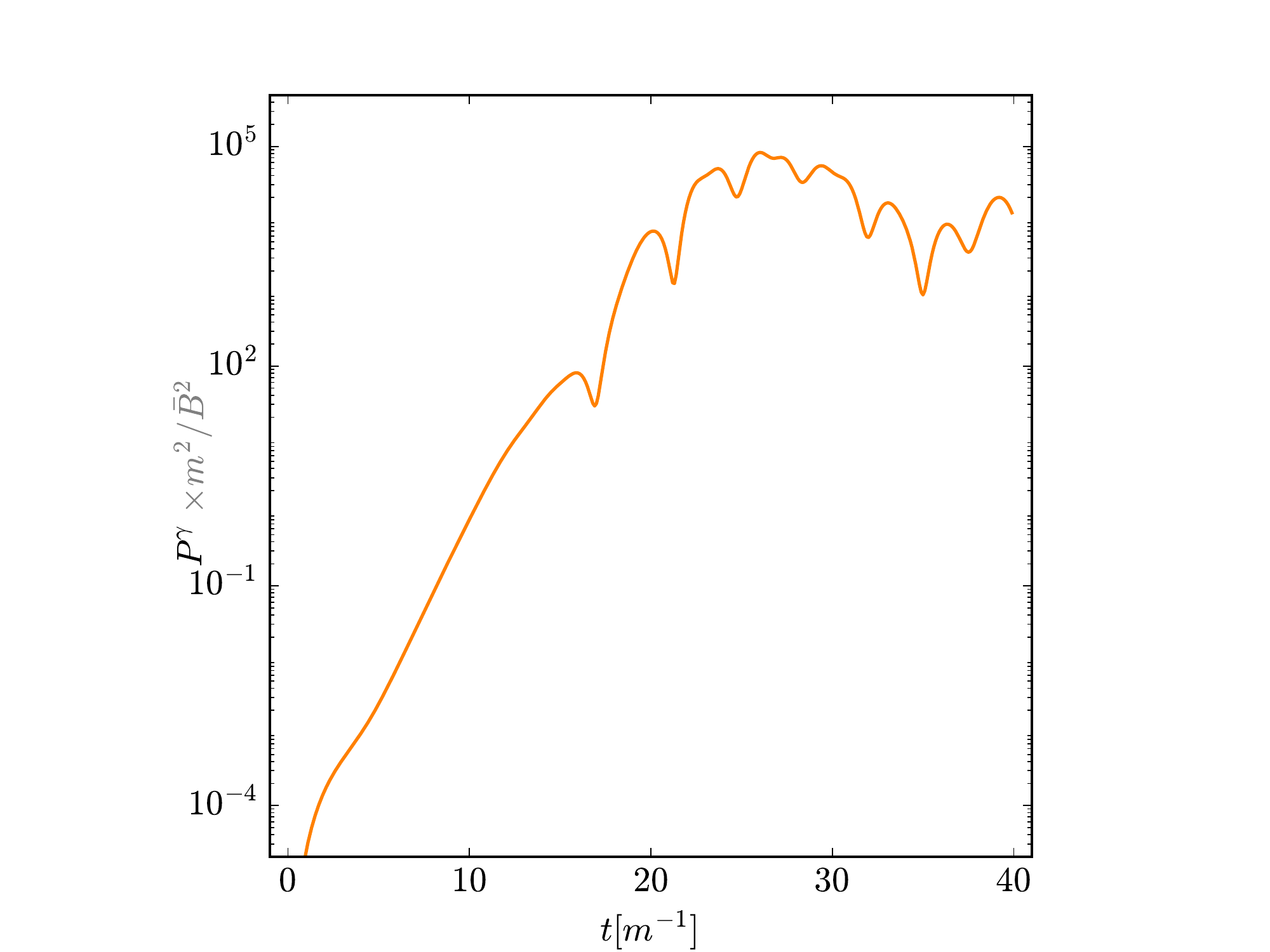}
&
\hspace{-1cm}
\includegraphics[width=0.55\textwidth,trim=2cm 0cm 2.5cm 1cm, clip]{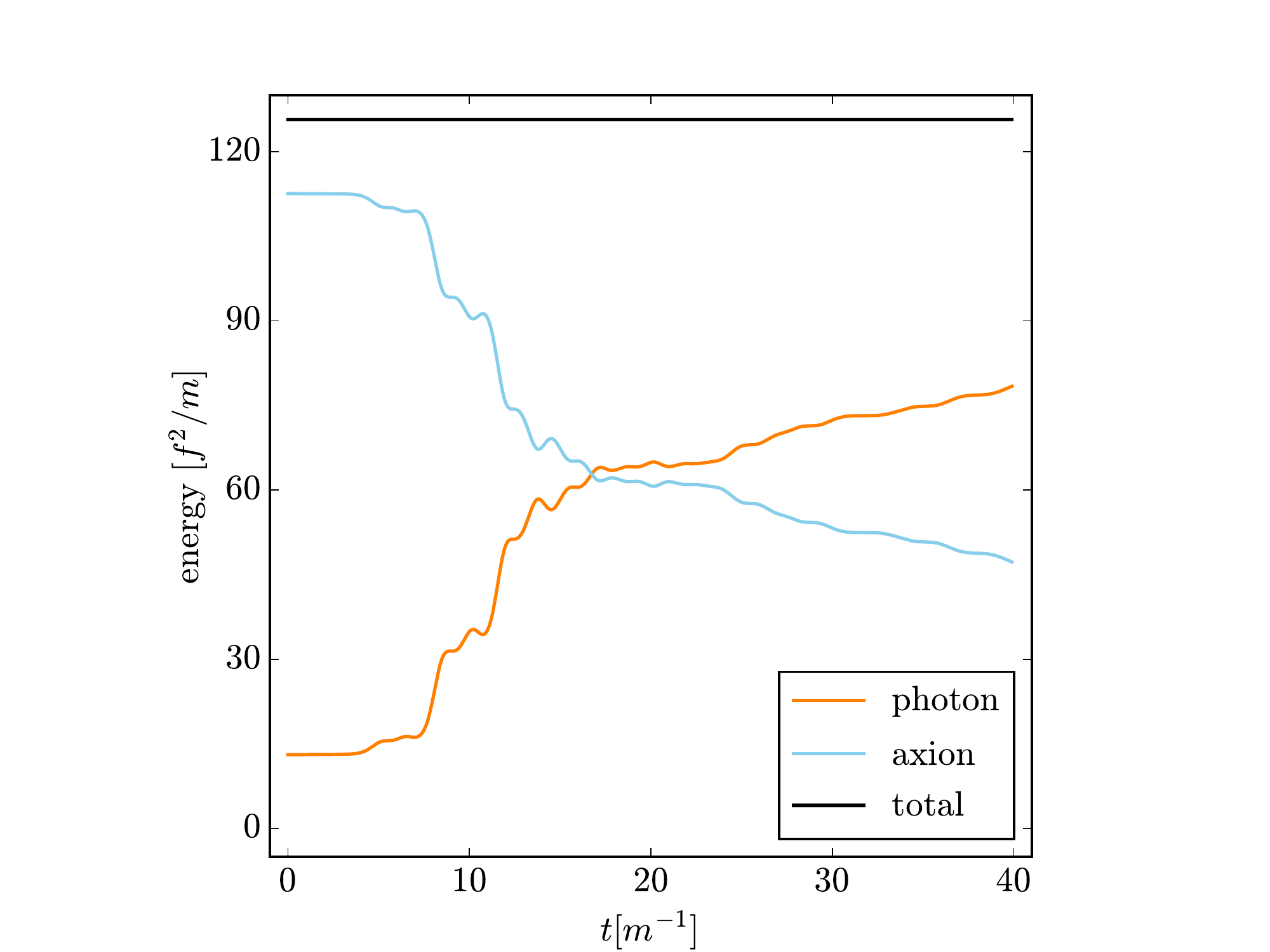}
\end{tabular}
\end{center}
\caption{
Power radiated by a dense soliton for large effective coupling $\Cc=1.7$ (left panel). In this regime the power grows exponentially with time, however backreaction eventually curtails this growth when the radiated electromagnetic energy becomes comparable to the initial energy of the soliton (right panel).
}
\label{fig:backreaction}
\end{figure}

To include this backreaction on the soliton, we evolve the fully coupled axion-photon system dynamically. The results are shown in Fig.~\ref{fig:backreaction}. The key point to note is that backreaction naturally regulates the exponentially growing radiated power once sufficient energy has been extracted from the soliton.

\section{Medium effects and coherence}
\label{sec:medium_effects}

In order to illuminate the connection between our calculation and calculations of axion-photon conversion in the literature, we address three issues here.  
First we discuss how our calculation should be modified in the presence of a medium, such as the dense plasma around a compact star, and the associated phenomenon of resonant conversion.  
Second we clarify how the axion star's coherence has affected the final radiation power.  
Third we compare our calculation with a perturbative calculation of the axion-photon conversion probability.  Here, we will restrict our attention to leading order in $\g\varphi_0$ to simplify the discussion.

\subsection{Medium effects and ``resonant" conversion}\label{sub:medium}
Our previous calculation was done with the axion star in the presence of background electromagnetic fields, with no other medium present at the background level. However, in applications to astrophysical scenarios, such as axion-stars in the magnetosphere of a neutron star, a plasma is present that leads to the photon having an effective mass $\omega_p$. The effect of such a constant effective mass $\omega_p<\omega$ can be approximately taken into account by modifying eq. \eqref{eq:Ewave1} and \eqref{eq:Bwave1} making the replacement $\nabla^2\rightarrow \left({\nabla}^2-\omega_p^2\right)$. Note that if $\omega<\omega_p$ the propagating mode  would be exponentially suppressed.\footnote{Since we assume the background $E$ and $B$ fields to be constant, it is natural to assume $\omega_p$ is constant, although in practice it does depend on the spatially varying free charge density  also.  In neutron star atmospheres, an approximation to the plasma frequency is given by $\omega_p=\sqrt{4\pi\alpha_{\rm em} n_e/m_e}$ where $n_e$ is the Goldreich-Julian charge density \cite{Goldreich:1969sb}, and $m_e$ is the mass of the electron. Note that $n_e(\bm{x})\sim \bm{\Omega}\cdot\bm{B}(\bm{x})$ where $\Omega$ is the angular velocity of the neutron star.}

\begin{figure}[t]
\begin{center}
\includegraphics[width=0.6\textwidth]{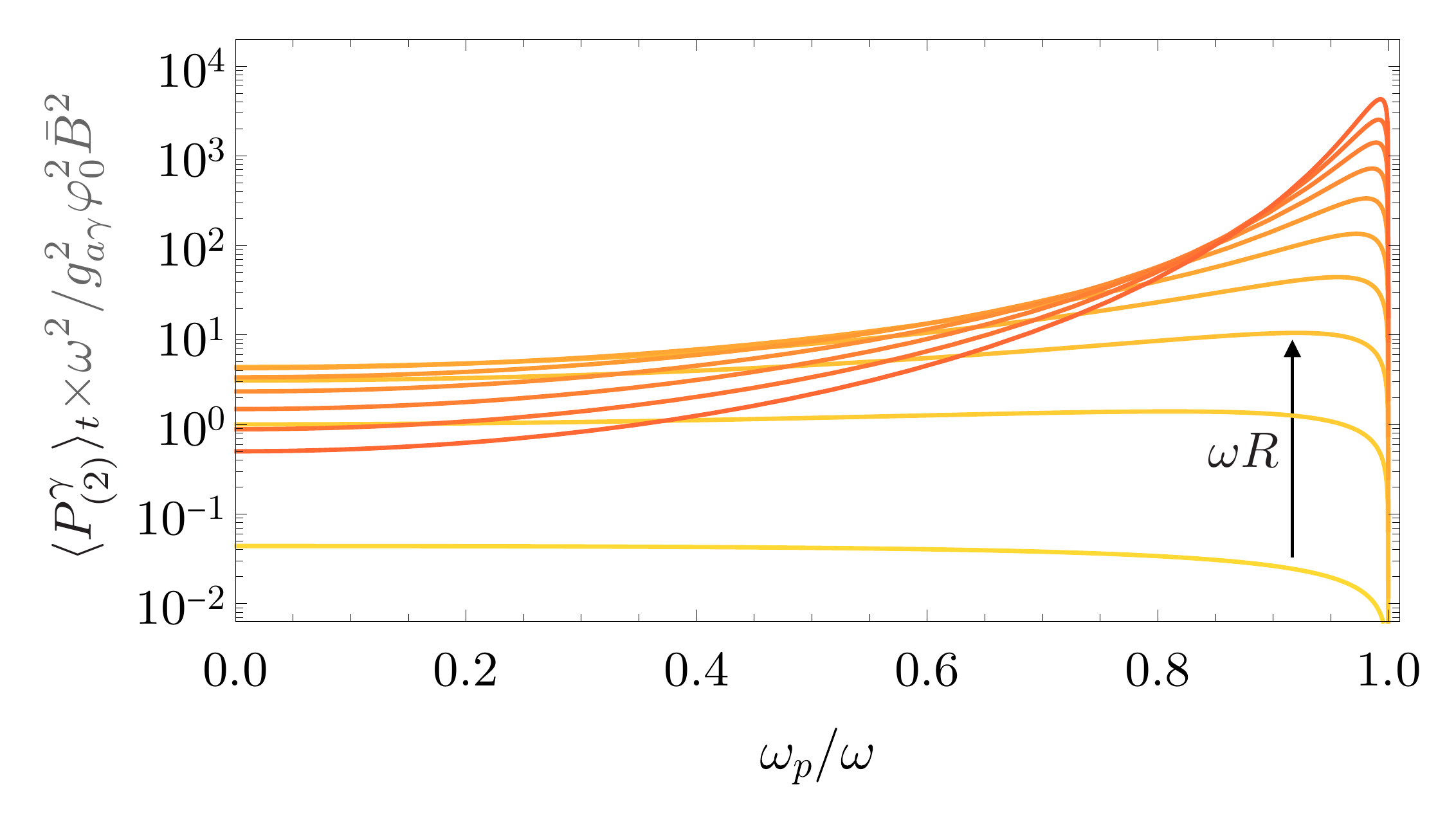}
\end{center}
\caption{
    The dependence of the time averaged power radiated by the axion star as a function of the plasma frequency $\omega_p$ and radius $R$ of the star. Note the ``resonant conversion" when $\omega_p\rightarrow\omega$ (the soliton frequency), and the strong dependence on the radius of the star. The plot should be understood under the assumption that the plasma frequency is constant within the radius of the star. In the above plot, $\omega R$ varies between $1$ and $10$, with the peak scaling as $(\omega R)^6$. Also note that while the power increases with radius for $\omega_p\approx \omega$, it decreases with radius when $\omega_p\ll \omega$  for large radii.}
\label{fig:ResonantConversion}
\end{figure}

Following through with the same calculation as before, but carefully keeping track of $\kappa$ and $\omega$ separately, we arrive at the generalization of our equation for the radiated power \eqref{eq:dipole1}:
\begin{align}
\label{eq:dipoleGenM}
&\langle P^\gamma_{(2)}\rangle_t =
\frac{\g^2\omega^4}{12\pi}\frac{\kappa}{\omega}\tilde{\varphi}^2(\kappa)\left(\bar{\Bvec}^2+\bar{\Evec}^2\right),\qquad\textrm{where}\qquad{\kappa}\equiv\sqrt{\omega^2-\omega_p^2}\,,
\end{align}
where $\tilde{\varphi}(\kappa)$ is the Fourier transform of the axion field profile at $|\kvec|=\kappa$. 
Note that for $\omega_p=0$, $\kappa=\omega$ and we recover our earlier result without the medium (\ref{eq:dipole1}). However, for $\omega_p\rightarrow \omega$ (``resonant conversion" domain), we have to be careful. Importantly, in the limit $\kappa\rightarrow 0$, we can remove the exponential suppression in $\tilde{\varphi}(\kappa)$ (see eq.~\eqref{eq:profileF}), and we obtain
\begin{align}
\label{eq:dipoleGenMApprox}
&\langle P^\gamma_{(2)}\rangle_t =
\frac{\pi (\g\varphi_0)^2}{48\omega^2}\left(\pi \omega R\right)^6 \sqrt{1-\frac{\omega_p^2}{\omega^2}}\left(\bar{\Bvec}^2+\bar{\Evec}^2\right)+\mathcal{O}\left[\left(1-\omega_p^2/\omega^2\right)^{3/2}\right]\,.
\end{align}
In the above expression we have assumed that $\omega_p$ does not vary within $R$. Notice that the medium effects point us to move to larger axion stars to get a large amount of power emitted, whereas without the medium, we must limit ourselves to a smaller radius because of the exponential suppression (assuming $\omega_p$ is constant on the scale $R$).

To see the detailed dependence of the radiated power on the radius $R$ and $\omega_p$, see Fig.~\ref{fig:ResonantConversion}. Note the large enhancement as we increase the radius for $\omega_p$ approaching $\omega$ from below. This is the resonant conversion. The ratio of the power emitted when $\omega_p\approx \omega$ above, compared to that when $\omega_p=0$ is $(1/16)(\pi \omega R)^2 e^{\pi\omega R}\sqrt{1-\omega_p^2/\omega^2}$. 

\subsection{Spatio-temporal coherence}\label{sub:coherence}
The dipole radiation power that we calculated in Sec.~\ref{sec:dipole} was derived with the assumption that the value of the axion field at different points across the soliton all oscillate in phase.  That is, we have phase coherence across the entire configuration. It is worth exploring the importance of this coherence for our results. To this end, suppose that the charge density generated by our axion configuration is replaced by $N$ idealized, equally spaced charge dipoles. Each dipole oscillates with frequency $\omega$ but a random phase $\theta_j$. The total charge density is 

\begin{figure}[t]
\begin{center}
\includegraphics[width=0.8\textwidth]{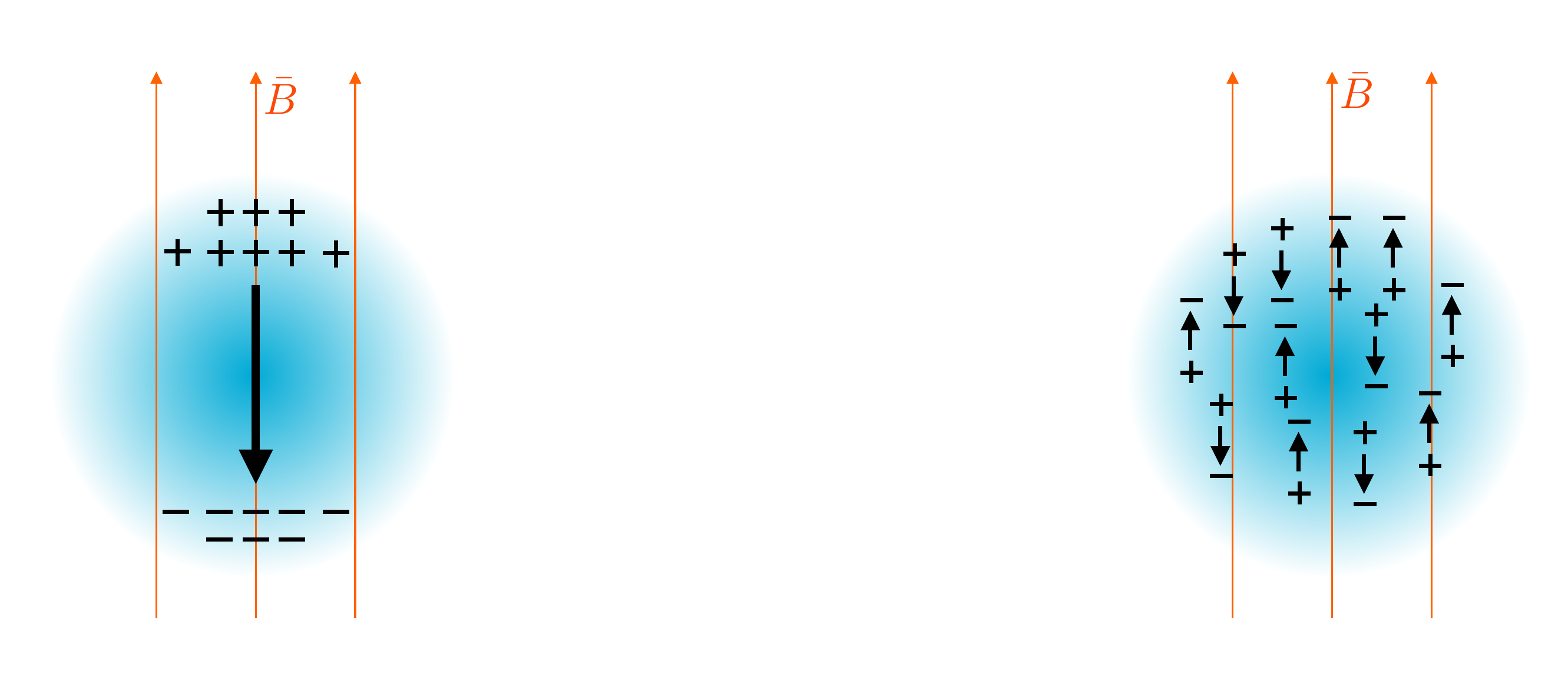}
\end{center}
\caption{A soliton of the axion field in an external magnetic field creates a coherently oscillating dipole configuration, which leads to dipole electromagnetic radiation. If instead, we replace the solition with $N$ oscillating dipoles with random phases, we can get less power radiated than the coherent soliton case for sufficiently large $N$. }
\label{fig:coherence}
\end{figure}

\Beq
\varrho_{(1)}^N(\xvec) =\sum_{j=1}^N \, (\Delta x)^3 \varrho_{(1)}(\xvec_j) \, \delta(\xvec-\xvec_j) \, e^{i\theta_j}\,,
\Eeq
as shown in Fig. \ref{fig:coherence}. Recall that $\varrho_{(1)}(\bm{x})=-\g \nabla\varphi\cdot\bar{\bm{B}}$. Then, under the assumption that the phases are random, we have 
\Beq
|\tilde{\varrho}_{(1)}^N(\bm{k})|^2=\left|\sum_{i=1}^N(\Delta x)^3\varrho_{(1)}(\xvec_j)e^{i\bm{k}\cdot\xvec_j+i\theta_j}\right|^2\sim \frac{V}{ N}\sum_{i=1}^N(\Delta x)^3\left|\varrho_{(1)}(\xvec_j)\right|^2\rightarrow \frac{Q_0^2}{N}\,,
\Eeq
where $\bm{k}=\omega \xhat$, we have assumed that $N$ is large, $V$ is the volume in which the $\varrho_{(1)}(\xvec)$ is non-zero. Note that $(\Delta x)^3=V/N$, and we defined 
\Beq
Q_0\equiv \sqrt{V\int \! \ud^3 \xvec \left|\varrho_{(1)}(\xvec)\right|^2}\,.
\Eeq
Since the power radiated is proportional to $|\varrho^N_{(1)}(\bm{k})|^2$, we can now compare the coherent and incoherent cases
\Beq
\frac{\langle P^{\gamma}_{(2)N}\rangle_t}{\langle P^\gamma_{(2)}\rangle_t}=\frac{|\tilde{\varrho}_{(1)}^N(\bm{k})|^2}{|\tilde{\varrho}_{(1)}(\bm{k})|^2}\sim \frac{Q_0^2}{N|\tilde{\varrho}_{(1)}(\bm{k})|^2}\,.
\Eeq
That is, if $N>Q_0^2|\tilde{\varrho}_{(1)}(\bm{k})|^{-2}$, the radiated power will be larger from a coherent configuration. For our $\sech$ profile and a constant background $B$ field, we have $|\tilde{\varrho}_{(1)}(\bm{k})|^2\approx (4\pi^2\bar{B}^2/\omega^4)(\g\varphi_0)^2(\pi\omega R)^4e^{-\pi\omega R}$ and $Q_0^2\sim 10^{-1}(\pi\omega R)^4 (\g\varphi_0)^2\bar{B}^2/\omega^4 $, which tells us that we need $N\gtrsim 10^{-3}e^{\pi \omega R}$ for coherence to win.\footnote{There can be purely numerical coefficients in front that depend on choice of $V$ and the details of the profile. The  scaling with $\omega R$, is the main result we want to focus on.}

For some localized configuration of radius $R$ with a characteristic density $\varrho_0\sim Q_0/R^3$, we can define a coherence length:
\Beq
\lambda_C\equiv R\left(\frac{|\tilde{\varrho}_{(1)}(\bm{k})|}{\varrho_0 R^3}\right)^{2/3}\,.
\Eeq
If we subdivide the volume of our coherent configuration into $N$ incoherent regions, each with a volume smaller than $\lambda_C^3$, then the power radiated from the coherent configuration will be larger. For our specific case of interest related to our soliton profile, we get $\lambda_C\sim e^{-\pi\omega R/3}R$. Hence for large radius configurations, incoherent emission will typically dominate over the coherent one.

\subsection{Axion-photon conversion probability}

We have calculated the electromagnetic radiation from an axion star in an external magnetic field by recognizing that the soliton behaves like a coherent dipole antenna.  
However, the conversion of axions into photons can be understood from a different point of view.  
The axion and photon fields mix with one another in the presence of an external magnetic field~\cite{Raffelt:1987im}, and an incident axion develops a nonzero probability to be detected later as a photon.  
This phenomenon is the basis of many laboratory probes of axions~\cite{Sikivie:1983ip}.  

In the presence of an external magnetic field $\bar{\bm{B}}$, the action from eq.~\eqref{eq:action_c} contains a mixing, $\mathcal{L}_\mathrm{mix} = - g_{a\gamma} \phi \dot{\bm{A}} \cdot \bar{\bm{B}}$, where we work in the Weyl gauge with $A^0 = 0$.  
To leading order in the coupling, the probability for an axion with momentum $p^\mu = \{E_p, {\bm p} \}$ to convert into a photon with momentum $k^\mu = \{\omega_k, {\bm k} \}$ is~\cite{Ioannisian:2017srr} 
\begin{align}
    \mathbb{P}_{a\to\gamma} = \frac{g_{a\gamma}^2}{4} \, \Bigl( \bigl| \tilde{\bm B}(q_+) \bigr|^2 + \bigl| \tilde{\bm B}(q_-) \bigr|^2 \Bigr)\,, 
\end{align}
where $\tilde{\bm{B}}(q_z)$ is the Fourier transform of $\bar{\bm{B}}(z)$.  
We have assumed that the magnetic field is static and only varies in the $z$ direction.  
The longitudinal momentum transfer, $q_z = k_z - p_z$, is restricted by energy and transverse momentum conservation to only take on two values, $q_\pm = \pm \sqrt{p_z^2 + m^2} - p_z$. If the magnetic field has a top-hat shaped profile, meaning that it is only nonzero (and takes value $B_0$) for a region of longitudinal distance $\ell$, then the Fourier transform is $\tilde{\bm{B}}(q_z) = (2B_0/q_z) \, \sin(q_z \ell/2)$. Moreover if the incident axion is non-relativistic, then $q_\pm \approx m$.  For a fiducial volume $V$ containing $N_a$ axions, the power per unit area of photons being emitted in the $z$-direction is $P^\gamma/A = \omega_k N_a \mathbb{P}_{a\to\gamma} / V$, which corresponds to $P^\gamma = \rho_a \mathbb{P}_{a\to\gamma} A$ where $\rho_a = m N_a / V \sim m^2 \varphi_0^2$ is the axion energy density and $\omega_k = m$.  

It is interesting to compare this calculation with the classical dipole radiation formula from eq.~\eqref{eq:dipoleB}.  
They display a similar parametric behavior, and both powers scale as $P^\gamma \propto g_{a\gamma}^2 \varphi_0^2 \bar{{B}}^2 R^2$.  
If we further identify the size of the B-field filling region with the radius of the axion star, $\ell = R$, then we have a factor of $\sin^2(mR)$, which also arises from an axion star with a top-hat density profile $\tilde{\varphi}^2(\omega=m) \propto \sin^2(mR)$.  
However, the two calculations are not necessarily equivalent, since eq.~\eqref{eq:dipoleB} is the power output from an (inhomogeneous) axion star in a homogeneous magnetic field, whereas the estimate above is for a homogeneous axion flux in an inhomogeneous (longitudinally-varying) magnetic field.  
We believe that these two approaches will yield consistent results when put onto the same footing (soliton structure and coherence), and we defer this investigation to future work.\footnote{Furthermore, at larger couplings , non-perturbative effects (Bose-effects) should be included in the framework of this calculation.}

\section{Observational signatures}
\label{sec:observability}

The central goal of our work is to understand the emission of electromagnetic radiation that occurs when an axion star passes through a strong electromagnetic field.  
We have seen that the radiation spectrum peaks at $E \sim \omega \sim m_a$, which corresponds to radio frequencies for typical axion masses.  
We have also seen that the radiation power grows as $P^\gamma \propto \bar{B}^2$ with the strength of the external magnetic field.  
In this section we will discuss how this phenomenon could lead to a variety of observational signatures in different environments with strong magnetic fields. We again restrict our attention to results at leading order in $\g$, although the large $\g$ results might lead to more radiated power in some cases.  

Using the dipole approximation from eq.~\eqref{eq:dipole1} and \eqref{eq:dipoleGenMApprox}, the luminosity ($L \equiv \langle P^\gamma_{(2)}\rangle_t$) of an axion star in a background magnetic field (with strength $\bar{B}$) is estimated as 
\begin{equation}\label{eq:power_master}
\begin{split}
    L & \simeq 
    \bigl( 4 \times 10^{22} \ \mathrm{W} \bigr) 
    \left(\frac{m}{10^{-5} \ \mathrm{eV}} \right)^{-2} 
    \left(\frac{g_{a\gamma}}{0.66 \times 10^{-10} \ \mathrm{GeV}^{-1}}\right)^2
    \left(\frac{f}{10^{10} \ \mathrm{GeV}}\right)^{-2}
    \\ & \qquad \times 
    \left(\frac{\bar{\bm{B}}}{10^{10} \ \mathrm{G}}\right)^2 
    \left(\frac{\varphi_0}{f}\right)^2 
    \mathcal{F}(\omega R,\omega_p/\omega),
    \;
\end{split}
\end{equation}
where we have normalized the axion-photon coupling $g_{a\gamma}$ to the 95\% CL upper limit from the CAST helioscope~\cite{Anastassopoulos:2017ftl}, and we have set $\omega = m$.  
We also remind the reader that $1 \, \mathrm{W} = 10^7 \, \mathrm{erg}/\mathrm{sec}$. The function $\mathcal{F}$ holds information about the soliton shape and plasma effects:
\begin{equation}\label{eq:Fcal_def}
\mathcal{F}(\omega R,\omega_p/\omega)\approx
  \begin{cases}
   (\pi \omega R)^4e^{-\pi \omega R}, & \text{for } \omega_p\approx 0\,, \\
    \dfrac{1}{16}(\pi\omega R)^6\sqrt{1-\omega_p^2/\omega^2}, & \text{for } \omega_p\approx \omega\,.
  \end{cases}
\end{equation}
where $\omega_p$ is the plasma frequency. Note the beneficial dependence on large radius and the lack of exponential suppression in the ``resonant" ($\omega_p\approx \omega)$ case. As long as the radius of the star is smaller than the size of the resonant region, our calculation holds, and leads to a large enhancement in the radiated power compared to the non-resonant case.

For the estimates in this section, we approximate the radiation spectrum as monochromatic, corresponding to a single spectral line.  
The frequency of this line is taken to be 
\begin{align}\label{eq:spectrum}
    \nu_\gamma = \frac{\omega}{2\pi} \approx \frac{m}{2\pi} \simeq \bigl( 2 \ \mathrm{GHz} \bigr) \left( \frac{m}{10^{-5} \ \mathrm{eV}} \right) 
    \;,
\end{align}
and we take the width of the line, i.e. the signal bandwidth, to be $\Delta \nu_\gamma \sim \nu_\gamma$.  For these fiducial parameters, we also note that the mass scale and radius of a very dense axion star (soliton) are expected to be on the order of 
\begin{equation}
\begin{split}
    M_\mathrm{sol} & \sim 10^2f^2 / m \simeq \bigl( 2 \times 10^9 \ \mathrm{kg} \bigr) \left( \frac{f}{10^{10} \ \mathrm{GeV}} \right)^{2} \left( \frac{m}{10^{-5} \ \mathrm{eV}} \right)^{-1} \\ 
    R_\mathrm{sol} & \sim 2m^{-1} \simeq \bigl( 4 \ \mathrm{cm} \bigr) \left( \frac{m}{10^{-5} \ \mathrm{eV}} \right)^{-1}
    \;.
\end{split}
\end{equation}
Note that $2 \times 10^9 \ \mathrm{kg} \approx  10^{-21} \ M_\odot$. 

\subsection{Compact stars}

The strongest magnetic fields in the universe today can be found in the magnetospheres of compact stars.  
The magnetic field strength at the surface of a white dwarf star is typically $10^{6-8} \ \mathrm{G}$~\cite{Ferrario_2015} whereas the smaller neutron stars can reach $10^{12-14} \ \mathrm{G}$~\cite{Olausen:2013bpa}.  
If an axion star were to encounter these extreme magnetic fields, the result would be a sudden and extreme release of electromagnetic radiation~\cite{Iwazaki:2014wka}.  

If the compact star is a distance $d_\star$ away, then the flux of radiation reaching Earth is $F = L / (4 \pi d_\star^2)$, which can be measured in $\mathrm{erg}/\mathrm{cm}^2/\mathrm{sec}$.  
The corresponding spectral flux density is calculated as $S = F/B$ where $B = \Delta \nu_\gamma = \omega / 2\pi$ is the signal bandwidth.  
For a nearby star, the spectral flux density evaluates to 
\begin{equation}\label{eq:S_compact_star}
\begin{split}
    S & \simeq 
    \bigl( 2 \times 10^{7} \ \mu\mathrm{Jy} \bigr) 
    \left(\frac{d_\star}{100 \ \mathrm{pc}} \right)^{-2} 
    \left(\frac{m}{10^{-5} \ \mathrm{eV}} \right)^{-3} 
    \left(\frac{g_{a\gamma}}{0.66 \times 10^{-10} \ \mathrm{GeV}^{-1}}\right)^2
    \\ & \qquad \times 
    \left(\frac{f}{10^{10} \ \mathrm{GeV}}\right)^{-2}
    \left(\frac{\bar{\bm{B}}}{10^{10} \ \mathrm{G}}\right)^2 
    \mathcal{F}(\omega R,\omega_p/\omega)
    \;,
\end{split}
\end{equation}
whereas the flux from a star at the galactic center ($d_\star \approx 8 \ \mathrm{kpc}$) would be reduced to $S \simeq 3 \times 10^{3} \ \mu\mathrm{Jy}$.  
For reference, an hour-long observation with a current or planned telescope (such as GBT, JVLA, or SKA) would have a flux sensitivity of $\delta S \sim 1 \ \mu\mathrm{Jy}$; see the estimates in Refs.~\cite{Bai:2017feq,Hook:2018iia}.  If an axion star were to pass through the magnetosphere of a compact star while it was being observed by a radio telescope, then the signal could be quite striking, even for modest couplings and field strengths. 

Since the compact star is surrounded by a plasma, this must be taken into account for the signal strength estimates.  
In Sec.~\ref{sub:medium} we have argued that the finite plasma density modifies the radiation spectrum, which is captured by $\mathcal{F}$ in eq.~\eqref{eq:Fcal_def}.  
This factor depends on the plasma frequency $\omega_p$, which is grows larger at points closer to the star, and $\mathcal{F}$ peaks near to where the plasma frequency matches the soliton's oscillation frequency, $\omega_p \approx \omega$, as shown in Fig.~\ref{fig:ResonantConversion}. 
For example, using the fiducial parameters in Ref.~\cite{Hook:2018iia}, the width of the resonance region is estimated to be $\omega L \sim \mathcal{O}(100)$.  
If the axion star's radius is $R \sim 0.1 \, L$ then Fig.~\ref{fig:ResonantConversion} implies an enhancement of $\mathcal{F} \sim 10^4$ to the spectral flux density estimate from eq.~\eqref{eq:S_compact_star}, which further increases the detectability.  

Even if an axion star's encounter with a compact star could be detected, we must address the expected rate of these encounters~\cite{Iwazaki:2014wka,Bai:2017feq,Buckley:2020fmh}.  
The encounter rate between a particular compact star and the ambient population of axion stars is estimated as $\Gamma = \sigma_\mathrm{eff} v_\mathrm{rel} n_\mathrm{as}$ where $\sigma_\mathrm{eff}$ is the effective cross sectional area for the scattering, $v_\mathrm{rel}$ is the typical relative velocity, and $n_\mathrm{as}$ is the number density of axion stars (near the target compact star).  
We can also write $n_\mathrm{as} = \rho_\mathrm{as} / M_\mathrm{sol}$ where $\rho_\mathrm{as}$ is the local mass density in axion stars and $M_\mathrm{sol}$ is the typical energy per axion star (soliton).  
The effective cross sectional area is further enhanced by the gravitational focusing factor, and we estimate $\sigma_\mathrm{eff} = (1 + v_\mathrm{esc}^2 / v_\mathrm{rel}^2) \pi R_\star^2$ where $v_\mathrm{esc}^2 =  M_\star / 4 \pi \mpl^2 R_\star$ is the escape velocity at the surface of the neutron star.  
Combining these factors allows us to estimate the encounter rate of axion stars with a particular white dwarf star to be 
\Beq\label{eq:Gamma}
    \Gamma \simeq 
    \bigl( 4 \times 10^{-5} \ \mathrm{hr}^{-1} \bigr) 
    \left( \frac{M_\star}{1 \, M_\odot} \right) 
    \left( \frac{R_\star}{0.01 \, R_\odot} \right) 
    \left( \frac{\rho_\mathrm{as}}{0.3 \ \mathrm{GeV} / \mathrm{cm}^3} \right) 
    \left( \frac{M_\mathrm{sol}}{10^{9} \ \mathrm{kg}} \right)^{-1} 
    \left( \frac{v_\mathrm{rel}}{10^{-3}} \right)^{-1} 
    \;, 
\Eeq
whereas the rate for encountering a neutron star (with $R_\star = 10 \ \mathrm{km}$ and other fiducial parameters unchanged) is $\Gamma \simeq 5 \times 10^{-8} \ \mathrm{hr}^{-1}$.  
The fiducial axion star density is taken to equal the local dark matter energy density near Earth, $\rho_\mathrm{dm} = 0.3 \ \mathrm{GeV} / \mathrm{cm}^3$, although axion stars are not expected to compose an $\mathcal{O}(1)$ fraction of the total dark matter density, which is typically dominated by a diffuse population of axion particles.

The estimate in \eqref{eq:Gamma} appears very unfavorable.  
For the fiducial parameters we expect a particular white dwarf star to encounter an axion star approximately once every $3$ years (or once every $2000$ years for a neutron star).  
However, there are several reasons why the rate might be enhanced over these estimates.  
First, the rate increases for compact stars at the galactic center or within dark matter subhalos (where $\rho_\mathrm{as}$ is higher if it tracks the dark matter density).  
Second we estimated $\sigma_\mathrm{eff}$ using the star's geometrical cross section, $\sim R_\star^2$, whereas the magnetic field extends far beyond the boundary of the star and scales like $B \sim r^{-3}$ for a magnetic dipole.  
Third, depending on the nature of the observation, it may be necessary to integrate over a finite region of the sky, such as toward the galactic center, which could contain many neutron stars, further increasing the encounter rate~\cite{Safdi:2018oeu}.  
Fourth, the fiducial axion star mass $M_\mathrm{sol} = 10^9 \ \mathrm{kg}$ is a free parameter, and a smaller value implies a larger encounter rate.  

\subsection{Direct detection in our solar system}

The phenomenon of electromagnetic radiation from an axion star in an external magnetic field could be used to develop a strategy for detecting axion stars when they encounter our solar system.  
The strongest magnetic fields generated in laboratories on Earth can reach strengths of a few Tesla, corresponding to $\sim 10^4 \ \mathrm{G}$.  
However, the flux of axion stars at Earth is expected to be quite low, making these signals very unlikely to be observed.  
The flux is estimated as $\Phi = \rho_\mathrm{as} v_\mathrm{rel} / M_\mathrm{sol}$, and the expected encounter rate with a 1 meter-scale detector is $\Gamma = \Phi (100 \ \mathrm{cm})^2 \sim (10^{-17} \ \mathrm{yr}^{-1}) \ (\rho_\mathrm{as} / 0.3 \ \mathrm{GeV} / \mathrm{cm}^3) (v_\mathrm{rel} / 10^{-3}) (M_\mathrm{sol} / 10^{9} \ \mathrm{kg})^{-1}$.  
Going beyond the confines of the laboratory, the Earth sustains its own magnetic field with a strength of $\sim 1 \ \mathrm{G}$.  
The smaller field strength would lead to a weaker signal, but the larger volume implies an increased encounter rate, $\Gamma \sim (10^{-4} \ \mathrm{yr}^{-1}) (M_\mathrm{sol} / 10^{9} \ \mathrm{kg})^{-1}$.  Finally, axion star encounters with the Sun's $\sim 1 \ \mathrm{G}$ magnetosphere could also provide a channel for detection.  
The encounter rate is enhanced by the Sun's much larger surface area, giving $\Gamma \sim (10^0 \ \mathrm{yr}^{-1}) (M_\mathrm{sol} / 10^{9} \ \mathrm{kg})^{-1}$, but the radiation power is much weaker, approximately $P \sim 10^{-24} \ L_\odot$, making this signal undetectable for the fiducial parameters.  

\subsection{Galactic magnetic field}

The axion stars in our Milky Way galaxy are continuously exposed to its $\sim 10^{-6} \ \mathrm{G}$ magnetic field.  
The corresponding electromagnetic radiation power is estimated using eq.~\eqref{eq:power_master}.  
For the fiducial parameters used above we find $P \sim 4 \times 10^{-10} \ \mathrm{W}$ for a single axion star.  
This power output is incredibly weak.  
For reference, if we sum the power output from all of the axion stars in a galaxy like the Milky Way (assuming that they make up all the dark matter), then the net power output is still only $10^{-3} \ L_\odot$!  However, see also Ref.~\cite{Moroi:2018vci} for a discussion of resonant axion-photon conversion in the intergalactic magnetic field.  

\subsection{Early universe}

We know very little about the extreme environment of the Universe during the first fractions of a second after the Big Bang.  Some theories predict that a magnetic field may have arisen during the period of cosmological inflation, post-inflationary reheating, or during a subsequent cosmological phase transition~\cite{Durrer:2013pga,Lozanov:2016pac}.  The strength of this primordial magnetic field may have been incredibly large by our every-day standards.  For instance a study of magnetogenesis from axion inflation~\cite{Adshead:2016iae} concluded that magnetic field generation could be so efficient as to transfer an $\mathcal{O}(1)$ fraction of the inflaton's energy into the magnetic field, leading to field strengths as large as $\sim 10^{52} \ \mathrm{G}$ at the end of inflation (for an inflaton mass of $m_\mathrm{inf} \sim 10^{14} \ \mathrm{GeV}$). 

Formation of oscillons and dense axion-star configurations has been explored in earlier works \cite{Kolb:1993hw,Amin:2011hj,Gleiser:2011xj,Lozanov:2017hjm,Hong:2017ooe,Muia:2019coe}. In well-motivated, observationally constrained models of inflation, the universe can become dominated by such solitons at the end of inflation (if the coupling to other fields is sufficiently weak) \cite{Amin:2011hj}. Similar phenomena are possible in moduli fields and other (pseudo-)scalars in the early universe. Typically, such configurations are long-lived compared to the age of the universe {\it then}, although they are not expected to survive until the present day. 

If the early universe were to contain both a strong primordial magnetic field and a population of axion stars~\cite{Yanagida:1987nf,Long:2015cza,Higaki:2013qka,Evoli:2016zhj,Dvornikov:2020hft}, then their interaction will induce electromagnetic radiation from the axion stars, thereby precipitating their decay (but also raise the question of whether the solitons would form in the first place).
Recall from the estimates in eq.~\eqref{eq:tbr} that an axion star with mass $M_\mathrm{sol} \sim 100 f^2/m$ emitting with a power $P^\gamma \sim 10 \bar{B}^2 (fg_{a\gamma})^2 / m^2$ would exhaust an $\mathcal{O}(1)$ fraction of its energy on a time scale of $\tau \sim M_\mathrm{sol} / P^\gamma \sim 10 m / g_{a\gamma}^2 \bar{B}^2$.  
Since $\bar{B}$ can be very large in the early universe, this axion star lifetime can potentially drop below the Hubble time scale at that time, which is $t_H \sim \mpl / T^2$ during radiation domination at temperature $T$.  We also note that even without strong magnetic fields, the solitons might be able to decay into photons rapidly due to collisions via mechanisms similar to those discussed in \cite{Amin:2020vja}. 

The phenomenon of magnetic-induced axion star decay would be challenging to test, since we have only a few handles on early universe physics.  
If the decay happens to occur during primordial nucleosynthesis, then the injection of electromagnetic radiation into the primordial plasma could potentially disrupt the formation of the light nuclei~\cite{Kawasaki:2017bqm}, and measurements of the light element abundances would provide an indirect constraint on this scenario.  
Nucleosynthesis also provides strong constraints on the QCD axion, even in the absence of a primordial magnetic field~\cite{Blum:2014vsa}. The interplay between solitons and electromagnetic fields can affect the rate of energy transfer and equation of state during reheating, as well as gravitational wave production, and spectral distortions during the early universe \cite{Lozanov:2019jxc,Lozanov:2017hjm,Antusch:2020iyq,Adshead:2015pva,Adshead:2019lbr,Kitajima:2018zco,Bolliet:2020ofj}.

\section{Summary and conclusion}
\label{sec:conclusion}

A spatially localized, periodically oscillating axion configuration (soliton: oscillon, axion star etc.) in background electromagnetic fields, sources electromagnetic radiation. We investigated this production analytically and numerically (with 3+1 dimensional lattice simulations when necessary), focusing in particular on the dependence of the emitted radiation on the characteristics of the axion field configuration as well as the strength of the coupling to the electromagnetic field. We also pointed out how the coherence of the soliton configuration, as well as the plasma effects, change the radiated energy in electromagnetic fields.

Our key results regarding the radiated power (luminosity) in electromagnetic waves are as follows:
\begin{itemize}

\item We delineated and verified the  boundary between bounded, constant luminosity solutions and exponentially growing ones based on axion-photon coupling and soliton properties. For a soliton with central amplitude $\varphi_0$, oscillation frequency $\omega$ and radius $R$, this boundary lies at $\Cc\equiv \g \varphi_0\omega R/4\sim 1$. This boundary is independent of the background electromagnetic fields.
    \item For $\Cc\ll1$, we get dipole radiation with a constant time-averaged luminosity. We derived an explicit formula for this dipole radiation, including an understanding of the strong (exponential) dependence on the radius of the solitons. 
    \item For the dense solitons (which we explore in detail in the numerics), we see a rich behavior of the radiated power as $\g$ is varied to explore all scenarios from $\Cc\ll1$ to $\Cc_{\rm crit}\sim 1$. Although the time-averaged radiated power remains constant in time, the details of the magnitude of the radiated power differ between background $E$ and $B$ field cases, and they are also sensitive to the details of the soliton configuration. For the $B$ field case, for all cases we have considered, we see a suppression compared to the dipole estimate and a non-monotonic behavior with $\Cc$. The same is not true for the $E$ field background.\footnote{We have checked that the radiated power is constant in this regime by doubling the linear size of the box and the duration of the simulation. While this constancy is expected from Floquet theory, it is not quite a proof since the possibility of band structure in $\Cc$ with complicated boundaries also exists which we might have missed out on numerically. Furthermore, the boundary at $\Cc=\Cc_{\rm crit}$ might be richer than just going from a constant-in-time radiated power to an exponentially growing one. We cannot exclude the possibility of a power-law behavior with time for the radiated power at this boundary. We leave this investigation to future work.} 
    
\item  For $\mathcal{C}\gtrsim 1$,  parametric resonance leads to an exponentially (in time) growing luminosity based on Floquet Theory. In the parametric resonance regime, background electromagnetic fields are unnecessary, small fluctuations can be sufficient. The exponential transfer of energy can be significant enough to cause backreaction on the soliton, and regulate photon production.

\item We explained the relevance of the coherently oscillating axion field configuration compared to an incoherent collection of dipoles and defined a critical coherence length which allows us to determine whether the coherent or incoherent configuration would radiate more efficiently. 

\item We explored how the presence of a plasma affects the radiation from a soliton. In particular, we find that when the plasma frequency is approximately equal to the oscillon frequency we get an enhanced resonant conversion to photons (`resonant' conversion, which is different from parametric resonance). 

\end{itemize}
There are a number of avenues for future work to extend our results. Our formalism and code includes background electric and magnetic fields together, however, we presented detailed numerical results for each separately. Considering them both together would introduce additional rich phenomenology which might be necessary, for example, when axion stars are boosted through static fields or when the astrophysical background fields themselves are time-dependent as is the case with  neutron stars.  In future work, we also plan to numerically include gravitational effects, such as tidal disruption, and take time-dependent medium effects around compact stars into account.

\section{Acknowledgements}
The numerical simulations were carried out on the NOTS cluster supported by the Center for Research Computing at Rice University. MA is supported by a NASA ATP theory grant NASA-ATP Grant No. 80NSSC20K0518, and PMS acknowledges support from STFC grant ST/P000703/1. We thank Yang Bai for helpful  comments on the draft. We also thank Kun Hu, Mudit Jain, Siyang Ling and Hongyi Zhang for useful discussions regarding the radiated power as well as neutron star atmospheres. 

\appendix
\section{Dipole radiation Green's function}
\label{sec:AppendixA}

In this appendix we solve the field equations, \eqref{eq:Ewave1} and \eqref{eq:Bwave1}, using the method of Green's functions, and we derive the time-averaged Poynting vector $\langle \Svec \rangle_t$.  
Consider the retarded Green's function
\begin{align}\label{eq:Greens_func}
    G(t, \xvec; \, t^\prime, \xvec^\prime)
    = 
    \int \! \! \frac{\ud^4k}{(2\pi)^4} \, 
    \frac{e^{i\kvec \cdot (\xvec-\xvec^\prime) - ik_0 (t - t^\prime)}}{(k_0+i\epsilon)^2-|\kvec|^2} 
    =
    -\frac{\delta(t - t^\prime -|\xvec-\xvec^\prime|)}{4\pi |\xvec-\xvec^\prime|} \ 
    \Theta(t-t^\prime)
    \;.
\end{align}
Solutions of eqs.~\eqref{eq:Ewave1} and \eqref{eq:Bwave1} are written as 
\begin{align}
    \Evec_{(1)}(t, \xvec) 
    & = 
    \int_{{\mathcal M}} \ud t^\prime \, \ud^3\xvec^\prime \ 
    G(t, \xvec; \, t^\prime, \xvec^\prime) \ 
    \left[ \dvec\rho_{(1)}(t^\prime, \xvec^\prime) + \dot{\Jvec}_{(1)}(t^\prime, \xvec^\prime) \right]
    \\ & \qquad 
    - \int_{\partial{\mathcal M}} \ud^3\xvec^\prime \
    G(t, \xvec; \, 0, \xvec^\prime) \ 
    \dot{\Evec}_{(1)}(0,\xvec^\prime) 
    \nonumber \\ & \qquad 
    + \int_{\partial{\mathcal M}} \ud^3\xvec^\prime \ 
    \partial_{t^\prime} G(t, \xvec; \, 0, \xvec^\prime) \ 
    \Evec_{(1)}(0,\xvec^\prime)
    , \nonumber \\
    \Bvec_{(1)}(t, \xvec) 
    & = 
    \int_{{\mathcal M}} \ud t^\prime \, \ud^3\xvec^\prime \ 
    G(t, \xvec; \, t^\prime, \xvec^\prime) \  \left[ -\dvec \times \Jvec_{(1)}(t^\prime, \xvec^\prime) \right]
    \\ & \qquad 
    - \int_{\partial{\mathcal M}} \ud^3\xvec^\prime \ 
    G(t, \xvec; \, 0, \xvec^\prime) \ 
    \dot{\Bvec}_{(1)}(0,\xvec^\prime)
    \nonumber \\ & \qquad 
    + \int_{\partial{\mathcal M}} \ud^3\xvec^\prime \ 
    \partial_{t^\prime} G(t, \xvec; \, 0, \xvec^\prime) \ 
    \Bvec_{(1)}(0,\xvec^\prime)
    \nonumber 
    \;,
\end{align}
where the integration contour ${\mathcal M}$ is understood as the upper half plane (time $t^\prime>0$), while the boundary $\partial {\mathcal M}$ is located at time $t^\prime=0$.  

In each expression above, the second and third terms enforce the initial conditions.  
However, for the purposes of calculating the late-time radiation at a point far away from the localized charge distribution, we can safely ignore these terms. Notice that these terms depend on the Green's function through $G(t, \xvec; \, 0, \xvec^\prime) \propto \delta(t - |\xvec - \xvec^\prime|)$.  
When $\xvec^\prime$ is restricted in the oscillon region and $\xvec$ is fixed for the observation, the delta function in the Green's function can not be satisfied given $t$ is big enough.  

At a point $\xvec$ that is far away from the localized charge distribution, we can approximate $|\xvec - \xvec^\prime| \approx |\xvec| - \xhat \cdot \xvec^\prime$ where $\xhat \equiv \xvec / |\xvec|$. Under this approximation, the fields are 
\begin{align}
    \Evec_{(1)}(t,\xvec) 
    & 
    \approx -{\rm Re} \left[ 
    \frac{e^{-i\omega t + i \omega |\xvec|}}{4\pi |\xvec|} 
    \int \! \ud^3\xvec^\prime \ 
    \bigl[ {\bm{\nabla}}\varrho_{(1)}(\xvec^\prime)-i\omega \jvec_{(1)}(\xvec^\prime) \bigr] \    
    e^{-i \omega \xhat \cdot \xvec^\prime} 
    \right]
    ,\\
    \Bvec_{(1)}(t,\xvec) 
    & 
    \approx -{\rm Re} \left[ 
    \frac{e^{-i\omega t + i \omega |\xvec|}}{4\pi |\xvec|} 
    \int \! \ud^3\xvec^\prime \ 
    \bigl[ -{\bm{\nabla}}\times\jvec_{(1)}(\xvec^\prime) \bigr] \ 
    e^{-i \omega \xhat \cdot \xvec^\prime} 
    \right]
    \;,
\end{align}
where we have used eq.~\eqref{eq:harmonicsource} and the Green's function from eq.~\eqref{eq:Greens_func} enforces $e^{-i \omega t^\prime} = e^{- i \omega t} e^{i \omega |\xvec - \xvec^\prime|} \approx e^{- i \omega t} e^{i \omega |\xvec|} e^{-i \omega \xhat \cdot \xvec^\prime}$.  
The integrals above are Fourier transforms, and we adopt the following conventions: 
\begin{align}
    f(\xvec) = \int \! \! \frac{\ud^3 \pvec}{(2\pi)^3} \, \tilde{f}(\pvec) \, e^{i\pvec\cdot\xvec}
    ,\quad
    \tilde{f}(\pvec) = \int \! \ud^3\xvec \, f(\xvec) \, e^{-i\pvec\cdot\xvec}
    \;.
\end{align}
This realization leads to
\begin{align}
    \Evec_{(1)}(t,\xvec) 
    &
    \approx  -{\rm Re}\left[ \frac{e^{-i \omega t + i \omega |\xvec|}}{4\pi |\xvec|} 
    \left[ i \kvec \tilde{\varrho}_{(1)}(\kvec) - i \omega \tilde{\jvec}_{(1)}(\kvec) \right] \right]
    ,\\
    \Bvec_{(1)}(t,\xvec) 
    &
    \approx -{\rm Re}\left[ \frac{e^{-i \omega t + i \omega |\xvec|}}{4\pi |\xvec|} 
    \left[-i \kvec \times \tilde{\jvec}_{(1)}(\kvec) \right] \right]
    ,
\label{eq:B}
\end{align}
with $\kvec \equiv \omega \xhat$.  
The Poynting vector is defined as 
\begin{align}
    \Svec_{(2)} = \Evec_{(1)} \times \Bvec_{(1)} 
    \;.
\end{align}
To calculate this quantity we use the arithmetic formula ${\rm Re}\left[a\right]{\rm Re}\left[b\right]={\rm Re}\left[ab^*+ab\right]/2$ and the current conservation equation $\omega  \tilde{\varrho}_{(1)}(\kvec) = \kvec \cdot \tilde{\jvec}_{(1)}(\kvec)$.  
Explicitly,
\begin{align}
    \Svec_{(2)}(t,\xvec) 
    = \frac{\omega \, \kvec}{32\pi^2 |\xvec|^2} 
    \Bigg( & 
    - |\tilde{\varrho}_{(1)}(\kvec)|^2
    + |\tilde{\jvec}_{(1)}(\kvec)|^2
    \\ & 
    -{\rm Re}\Big[
    e^{-i2\omega t} 
    e^{i2\omega |\xvec|} 
    \left(
    - \tilde{\varrho}_{(1)}^{2}(\kvec) 
    + \tilde{\jvec}_{(1)}^{2}(\kvec) 
    \right)
    \Big]
    \Bigg)\,.\nonumber 
\end{align}
The first two terms are independent of time $t$, while the third term oscillates with period $\pi/\omega$ and it vanishes upon taking the time average (over many oscillations cycles).  
Thus, the time-averaged Poynting vector is 
\begin{align}\label{eq:S2_final}
    \langle \Svec_{(2)}\rangle_t(\xvec) =
    \frac{\omega \, \kvec}{32\pi^2 |\xvec|^2} \ 
    \Big(
    -|\tilde{\varrho}_{(1)}(\kvec)|^2
    +|\tilde{\jvec}_{(1)}(\kvec)|^2
    \Big)
    \;,
\end{align}
where $\kvec = \omega \xhat$, for a spatially-localized, spherically-symmetric charge distribution that oscillates with period $2\pi/\omega$.  

A similar calculation goes through for a photon with nonzero mass $0 < m_\gamma < \omega$.  
In a medium $m_\gamma = \omega_p$ is the plasma frequency.  
The field equations, \eqref{eq:Ewave1} and \eqref{eq:Bwave1}, are extended to include the mass term.  
The Greens function in eq.~\eqref{eq:Greens_func} involves a massive propagator that enforces the dispersion relation, $k_0^2 - |\kvec|^2 = \omega_\mathrm{p}^2$.  
Ultimately the Poynting vector is expressed as in eq.~\eqref{eq:S2_final} but with the wavevector $\kvec = \omega \xhat$ replaced by $\kvec = \kappa \xhat$ with $\kappa = [\omega^2 - \omega_p^2]^{1/2}$.  
For larger photon masses, $\omega < \omega_p$, the fields are exponentially damped.  

\bibliographystyle{utphys}
\bibliography{reference}

\end{document}